\documentclass{aa}  
\usepackage{graphicx}
\usepackage{amsmath}
\usepackage[varg]{txfonts}
\usepackage{natbib}
\bibpunct{(}{)}{;}{a}{}{,}
\begin{document}
\title{Superbubble evolution in disk galaxies}
\subtitle{I. Study of blow-out by analytical models}


\author{V. Baumgartner
          \inst{1}
          \and
          D. Breitschwerdt\inst{2}
          }

\institute{
Institut f\"ur Astrophysik, Universit\"at Wien, T\"urkenschanzstr. 17, A-1180 Vienna, Austria\\
\email{verena.baumgartner@univie.ac.at}
\and
Zentrum f\"ur Astronomie und Astrophysik, Technische Universit\"at Berlin, Hardenbergstr. 36, D-10623 Berlin, Germany
             }

\date{Received 8 February 2013 / Accepted 1 August 2013}

 
\abstract
    {Galactic winds are a common phenomenon in starburst galaxies in the local universe as well as at higher 
redshifts. Their sources are superbubbles driven by 
sequential supernova explosions in star forming regions, which carve out large holes in the interstellar medium and 
    eject hot, metal enriched gas into the halo and to the galactic neighborhood.    }
   {We investigate the evolution of superbubbles in exponentially stratified disks.
We present advanced analytical models for the expansion of such bubbles and calculate their evolution in space and time. With 
these models one can derive the energy input 
that is needed for blow-out of superbubbles into the halo and derive the break-up of the shell, since   
Rayleigh-Taylor instabilities develop soon after a bubble starts to accelerate into the halo. }
   {The approximation of Kompaneets is modified in order to calculate velocity and acceleration of a bubble analytically.
Our new model differs from earlier ones, because it presents for the first time an analytical calculation for the expansion of superbubbles in an 
exponential 
density distribution driven by a time-dependent energy input rate. The time-sequence of supernova explosions of OB-stars is 
modeled using their main sequence lifetime and an initial mass function. }
   {We calculate the morphology and kinematics of superbubbles powered by three different kinds of energy input and we derive the energy input required
for blow-out as a function of the density and the scale height of the ambient interstellar medium.
The Rayleigh-Taylor instability timescale in the shell is calculated in order to estimate when the shell starts to fragment and finally breaks up. 
Analytical models are a very efficient tool for comparison to observations, like e.g. the Local Bubble and the W4 bubble discussed in this paper, and also give insight into the dynamics of superbubble evolution.}
%
{}


\maketitle
%
%
\section{Introduction}
Most massive stars are born in OB-associations in coeval starbursts on timescales of less than 1-2 Myr \citep{m99}.
These associations can contain a few up to many thousand OB-stars, so-called super star clusters, but typically have 20-40 members \citep{mk87}.
Energy and mass are injected through strong stellar winds and subsequent supernova (SN) explosions of stars with masses above
8 M$_{\odot}$.
The emerging shock fronts sweep-up the ambient interstellar medium (ISM) and, as the energy input in form of SN-explosions continues,
superbubbles (SBs) are produced, which may reach dimensions of kiloparsec-size \citep[e.g.][]{tsm03}.
The swept-up ISM collapses early in the evolution of the SB into a cool, thin, 
and dense shell \citep{cmw75} present in HI and H${\alpha}$ observations.
The bubble interior contains hot ($> 10^6 \, $K), rarefied material, usually associated with extended diffuse X-ray emission \citep{sta05}. 
Due to the stratification of the ISM in disk galaxies, the superbubbles can accelerate along the density and pressure gradient and blow out into the halo, 
appearing as elongated structures. 
Examples of huge bubbles and supergiant shells are the Cygnus SB with a diameter of 450 pc in the Milky Way 
\citep{ccb80} and the Aquila supershell extending at least
550 pc into the Galactic halo \citep{m96}. 
Our solar system itself is embedded in an HI cavity with a size of a few hundred parsecs
called the Local Bubble (\citeauthor{l03} \citeyear{l03}), and is most likely generated by stellar explosions in a nearby moving group (\citeauthor{bb02} \citeyear{bb02}).
Also in external systems like in the LMC \citep{cm90}, in NGC 253 \citep{s06} and 
M101 \citep{ksh91} such bubbles, holes and shells are observed.\\
The acceleration of the shell promotes Rayleigh-Taylor instabilities and after it is fully fragmented, 
only the walls of the SB are observed. Through such a chimney the hot pressurized SN-ejecta can escape into the halo.
The walls may be subject to the gravitational instability and interstellar clouds can form again, which triggers star formation 
\citep[e.g.][]{mk87}. This is observed, for example, on the border of the Orion-Eridanus SB \citep{lc09}.\\
The knowledge of SB evolution is crucial for understanding the so-called disk-halo connection and it also gives us information about the chemical evolution of the galaxies, 
the enrichment of the intergalactic as well as the intracluster medium.
The thick extraplanar layer of ionized hydrogen seen in many galaxies 
has probably been blown out of the disk into the halo by photoionization of OB-stars and correlated SNe \citep{t06}.
With the high star formation rate of starburst galaxies, the energy released by massive bursts of star formation can even push the gas out of the 
galactic potential, forming a galactic wind. Outflow rates of $0.1-10 \, $M$_{\odot}$/yr are common for starburst driven outflows \citep{bvc07}.
%
If the hot and metal enriched material is brought to the surrounding intergalactic medium, it will mix after some time,
increasing its metallicity. Galactic winds are observed in nearby galaxies \citep[e.g.][]{dwh98}, as well as up to redshifts 
of $z \sim 5$ \citep[e.g.][]{sw07,dss02}, thus being an ubiquitous phenomenon in star forming galaxies.
If the energy input is not high enough, the gas will fall back onto the disk due to gravity, after loss of pressure support, forming a galactic fountain \citep{sf76,a00}.
%
In this case, the heavy elements released by SN-explosions are returned back to the ISM in the disk, presumably spread over a wider area, and future generations of stars will incorporate them. 

Spitoni et al. (2008), also using the Kompaneets approximation, 
have investigated the expansion of a superbubble, its subsequent fragmentation and also the ballistic motion of the fragments including a drag term, which describes the interaction between a cloud and the halo gas.
In addition, these authors
have analyzed the chemical enrichment of superbubble shells, 
and their subsequent fragmentation by Rayleigh-Taylor instabilities, in order to compare the [O/Fe]-ratios of these blobs to high velocity clouds (HVCs). They find that HVCs are not part of the galactic fountain, 
and even for intermediate velocity clouds (IVCs), which are in the observed range of velocities and heights from the galactic plane, the observed [O/Fe]-ratios can only be reproduced by unrealistically low initial disk abundances.
The chemical enrichment of the intergalactic medium will be the subject of a forthcoming paper.\\
This paper is structured as follows: Section 2 shows how the evolution of superbubbles can be described. 
In Sect.~3 we present the results of this work, which is mainly 
the expansion of a bubble in space and time and also the onset of Rayleigh-Taylor instabilities in the shell.
In Sect.~4 the models are used to analyze two Milky-Way superbubbles. A discussion follows in Sect.~5 and summary and conclusions are presented in Sect.~6. 
%
%
%
%
%
\section{Superbubble evolution}
%
%
\subsection{ISM stratification}
For a homogeneous ISM, the propagation of a shock front originating from an instantaneous release of energy was described by \citet{s46} and \citet{t50}, while the 
expansion of an interstellar bubble with continuous wind energy injection was studied by \citet{cmw75} and \citet{w77}. 
Yet, the description of SB evolution in such a uniform ambient medium is only valid for early stages of evolution.
The vertical gas density distribution has a major effect on the larger superbubbles which have sizes exceeding the thickness of a galactic disk.
On these scales, the ISM structure is far from homogeneous.
In a Milky-Way-type galaxy, the warm neutral HI is plane-stratified and can be described by a symmetric exponential atmosphere with respect to the midplane 
of a galaxy \citep{l84}. The scale height of this layer is $\sim$ 500 pc. 
The Reynolds layer of warm ionized gas has a scale height of $H = 1.5 \,$kpc \citep{r89} and the highly ionized gas of the hot ($10^6 - 10^7 \,$K) halo surrounds the 
galaxy extending to $\sim 20 \,$kpc \citep{bl07} with a scale height of $\sim 4.4 \,$kpc \citep{ssl97}.
%
In the disk, the cold neutral and molecular gas are found to have $H \sim 100 \,$pc.\\
The expansion of superbubbles in a stratified medium was studied since many decades, both analytically \citep[e.g.][]{mc99} and 
numerically \citep[][hereafter MM88]{cg74,ti86,mm88}.
For an analytic description of superbubbles, Kompaneets' approximation (1960) is a very good choice.
Although it involves several simplifications (e.g.~no magnetic field), it represents the physical processes involved very well \citep{pls07}.
%
%
%
%
\subsection{Blow-out phenomenon and fragmentation of the shell}
A focus of this paper is to analyze the blow-out phenomenon: in an exponential as well in an homogeneous ISM, 
a bubble decelerates first. But due to the negative density gradient of the exponentially 
stratified medium and the resulting encounter with very rarefied gas, the bubble starts to accelerate into the halo and even beyond, if the energy input is 
high enough, at a certain time in its evolution. 
MM88 call this process blow-out, which is also used by \citet{s85} and \citet{ft00}, but there, blow-out involves complete escape of the gas from the galaxy.
From a more phenomenological point of view, \citet{h90} distinguishes between breakthrough bubbles, which break out of the dense disk and are observed as shells and
holes 
in the ISM, whereas blow-out bubbles break through all gas layers and inject mass and metals into the halo. 
In our definition (see Sect.~2.3), a SB will blow out of a specific gas layer at the time, when the outer shock accelerates, and if the shock stays strong all the time.
In particular, we want to determine the energy input required for blow-out into the halo, and its dependence on ISM parameters (see Sect.~3.2.).\\
%
%
Shortly after the acceleration has started, Rayleigh-Taylor instabilities (RTIs) appear at the interface between
bubble shell and hot shocked bubble interior. As the amplitudes of the perturbations grow, 
finger-like structures develop at the interface and vorticity of the flow increases due to shear stresses.
%
Finally, in the fully non-linear phase of the instability, turbulent mixing of the two layers starts, and the shell will eventually break-up and fragment. 
An azimuthal magnetic field in the shell will limit the growth rate of the instability due to magnetic tension forces, but not for all wave modes of the instability. In essence, only modes above a critical wave number will become unstable, giving a lower limit to the size of blobs \citep{bfe00}.
This happens first at the top of the expanding bubble, where the acceleration is highest. 
The clumps generated this way and the hot gas inside the bubble -- including the highly enriched material -- are expelled into the
halo of the galaxy or even into intergalactic or intracluster space, contributing to the chemical enrichment of galactic halo or intracluster 
medium.\\
After deriving the acceleration of the outer shock in Sect.~3.2, we analyze the timescales for the development of RTIs in the shell in Sect.~3.3 .
%
%
%
\subsection{Modeling superbubbles}
Originally developed to describe the propagation of a blastwave due to a strong nuclear explosion in the Earth's atmosphere, the approximation found by 
Kompaneets (1960) is also applicable to investigate the evolution of a superbubble in a disk galaxy analytically. 
We modify Kompaneets' approximation (KA) in order to describe not only a bubble driven by 
the energy deposited in a single explosion or as a continuous wind \citep[e.g.][]{s85}, but to produce analytical models for the expansion using a 
time-dependent energy input rate due to sequential SN-explosions of massive stars according to a galactic initial mass function (IMF).
The axially symmetric problem is described in cylindrical coordinates $(r, \, z)$. In order to use the
KA for the investigation of SB evolution the following assumptions have to be made:
(i) the pressure of the shocked gas is spatially uniform, (ii) almost all of the swept-up gas behind the shock front is located in a thin shell, and (iii)
the outer shock has to be strong all over the evolution of the bubble.
%
%
Using the third assumption we get our blow-out condition: if the outer shock has a Mach number $M \geq 3$, i.e.~ the upstream velocity of the gas in the shock frame is at least 3 times the sound speed of the warm neutral ISM at the transition of deceleration to acceleration, then the bubble will blow-out into the halo and reach regions of higher galactic latitudes. 
\citet[][hereafter MMN89]{mmn89} find via comparisons to their numerical simulations that the KA can be even used after the instabilities in the shell set in, 
because the pressure inside the bubble is not released very quickly.  
If the condition is not fulfilled (i.e. no strong shock), the system is obviously not energetic enough, hence cannot be described by the KA 
and will finally slow down and merge with the ISM like in the case of disk supernova remnants.
Blow-out usually occurs when the shell reaches between one \citep{vcb05} and three scale heights \citep{ft00}.
In the next section, this is confirmed and exact values for the height of the bubble are given, using different descriptions of the ambient ISM and
different ways of energy input (see Table \ref{accel}).
Compared to other groups using a dimensionless dynamical parameter introduced by MM88 to decide if blow-out will happen, our 
criterion's advantage is that the result is given in explicit numbers, and hence easier to use when compared to observations.
Additionally, we investigate if the acceleration of the bubble at the time where fragmentation occurs will exceed the gravitational acceleration near the galactic plane. Using this simple comparison we can 
ensure that fragments of the bubble and the hot bubble interior will be expelled into the halo instead of falling back onto the disk. \\ 
A pure exponential atmosphere was used in the KA, which was already modified for 
a radially stratified medium \citep{k92} and an inverse-square decreasing density \citep{kp98}. 
We adopted the original calculations to model the expansion of superbubbles in a more realistic fashion and investigate two cases of density distribution in our paper: 
the first one corresponds to a bubble evolving in an exponentially stratified medium symmetric to the midplane
\begin{equation}
\rho_I(z)=\rho_0 \cdot \exp[-\vert z \vert /H] \, ,
\label{erho1}
\end{equation}
where $\rho_0$ is the density in the midplane and $H$ is the scale height of the ISM. The cluster is located at $z=0.$
Since it is an idealization that OB-associations are only found in the galactic midplane, but are rather offset in z-direction, we examine in our second case 
an off-plane explosion, where the density law is given by
\begin{equation}
\rho_{II}(z)=\rho_1 \cdot \exp[-( z -z_0 ) /H]
\label{erho2}
\end{equation}
with $z_0$ being the center of the explosion. The density at $z=z_0$ is either derived from the relation $\rho_1= \rho_0 \cdot \exp[-z_0/H]$ or can be a known value.
Any real case encountered will be described by one of the two or lie in between, i.e. the offset is small enough for the midplane to be pierced by further explosions.
%
%
   \begin{figure}
   \centering
   \includegraphics[angle=270, width=5.5cm]{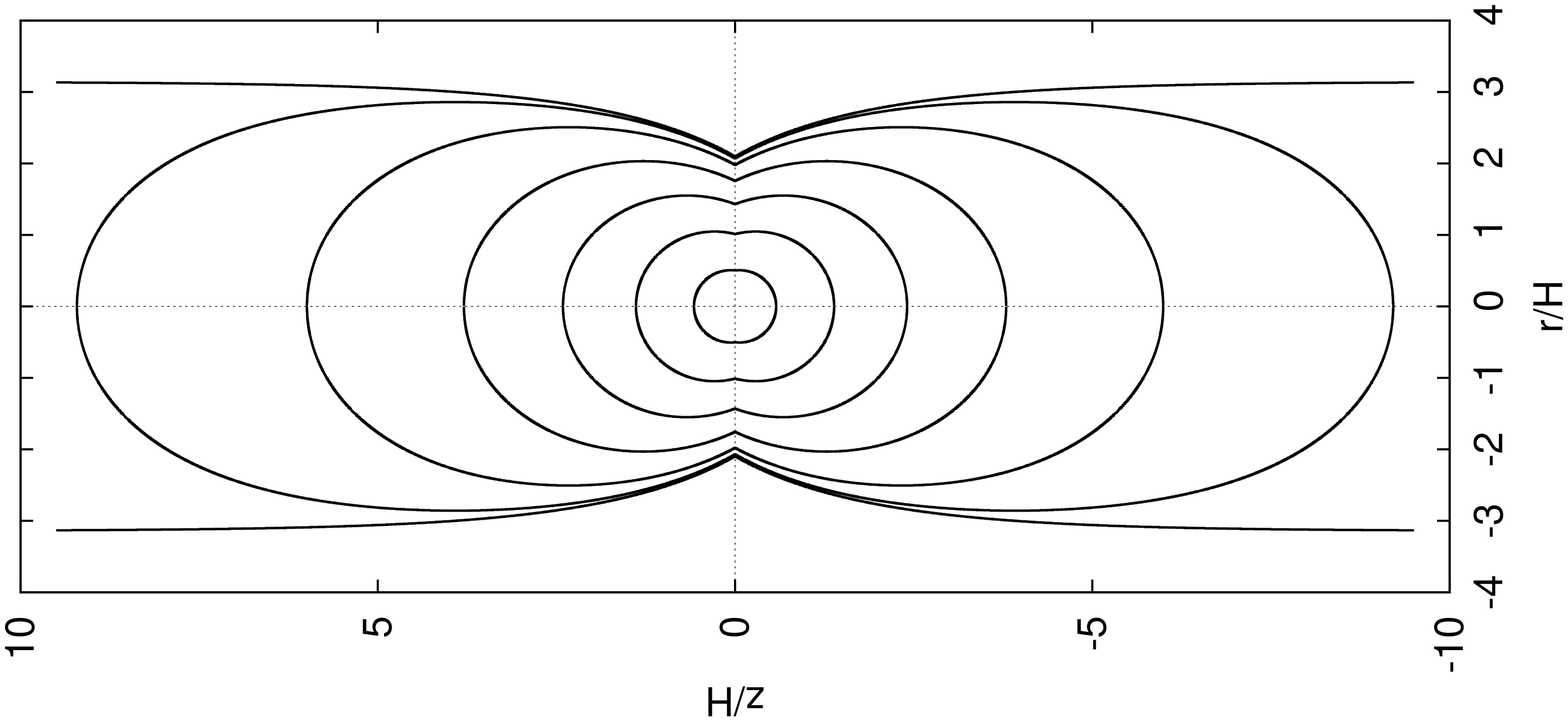}
   \caption{\label{SBsymm}Position of the shock front in the symmetric model at certain values of the dimensionless time variable 
$\tilde{y} = 0.5$, $1.0$, $1.4$, $1.7$, $1.9$, $1.98$, and $2.0$ with the energy source in the galactic midplane.}
   \end{figure}
%
   \begin{figure}
   \centering
   \includegraphics[angle=270, width=5.5cm]{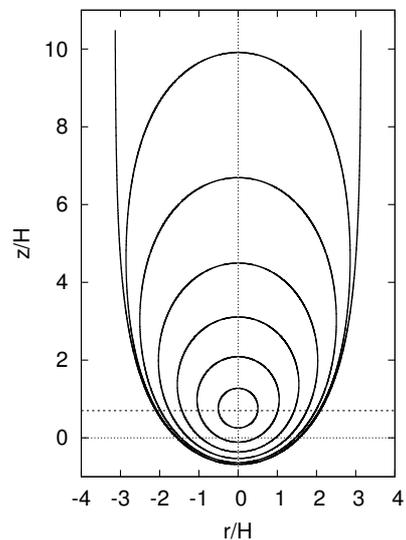}
   \caption{\label{SBoff} Same as Fig.~\ref{SBsymm}, but for the off-plane model. The energy source is located at $z_0=0.7\,$H above the plane (indicated by the dashed line).}
   \end{figure}

\noindent
To get a spatial solution the Rankine-Hugoniot conditions for a strong shock for every point of the azimuthally symmetric shock surface have to 
be solved \citep[e.g.][]{bs95}. 
This gives the normal component of the expansion speed
\begin{equation}
v_n = \sqrt{
\frac{\gamma + 1} {2} \cdot \frac{P(t)} {\rho \,(z)}
} \, , 
\label{evn}
\end{equation}
where $\gamma$ is the ratio of specific heats ($\gamma =5/3$ for a monatomic ideal gas).
We need to know the pressure $P(t)$ in the bubble, 
\begin{equation}
P(t) = (\gamma - 1) \, \frac {E_{\rm{th}}}{\Omega(t)}
\end{equation}
with $\Omega(t)$ being the volume confined by the shock and $E_{\rm{th}}$ the thermal energy in the hot shocked gas region. 
The volume is given by the integral
\begin{equation}
\Omega(t) = \pi \int_{z_2}^{z_1} r^2(z,t) \, dz \, .
\label{eomega}
\end{equation}
Introducing a transformed time variable (in units of a length)
\begin{equation}
y =  \int_{0}^{\, t} \sqrt{\frac{\gamma^2 - 1}{2}\frac{E_{\rm{th}}}{\rho_0 \cdot \Omega(t)}} \, dt
\label{ey}
\end{equation}
%
%
makes it easier to solve the PDE which is obtained after rearranging and equating Eq.~(\ref{evn}) and the time derivative of Eq.~(\ref{ey}) and by assuming that the shock front is a time-dependent surface, defined as $f(r,z,t) = 0$
\begin{equation}
\left( \frac{\partial r}{\partial y} \right)^2 - \frac{\rho_0}{\rho(z)} \, \left(1 + \left( \frac{\partial r}{\partial z} \right) ^2 \right)
= 0 \, .
\end{equation}
By separation of variables one gets the solution $r = r(y,\, z)$, which describes the evolution of the half width-extension of the bubble parallel to the 
galactic plane.
For the symmetric density law this is
\begin{equation}
r_I(y,z)=\left\{
\begin{array}{cl}
2H \, \arccos \left[ \frac{1}{2} e^{z/2H} \left(1 - \frac{y^2}{4H^2} + e^{-z/H}  \right) \right], \, z \geq 0\\[0.2cm]
2H \, \arccos \left[ \frac{1}{2} e^{-z/2H} \left( 1 - \frac{y^2}{4H^2} + e^{z/H}  \right) \right], \, z < 0 
\end{array}
\right.
\label{erI}
\end{equation}
and for an off-plane explosion we find 
\begin{equation}
r_{II}(y, z)=2H\arccos \left[ \frac{1}{2} e^{(z-z_0)/2H} \left(1 - \frac{y^2}{4H^2} + e^{-(z-z_0)/H} \right) \right].
\label{erII}
\end{equation}

\noindent
Due to the explicit relation between time $t$ and $y$, the parameter $y$ is used to represent the evolution of the bubble.
At this point we introduce a dimensionless parameter $\tilde{y} = \frac{y}{H}$.
Figs.~\ref{SBsymm} and \ref{SBoff} show the position of the outer shock at
different values of $\tilde{y}$ for the symmetric and off-plane model, respectively.\\
A few points of the bubble's surface are special, because explicit equations are available for them from the solution (Eqs.~(\ref{erI}) and (\ref{erII})).
When $r(\tilde{y},z) = 0$, the top $z_u$ and bottom $z_d$ of the bubble can be evaluated.
In the case of the symmetrically decreasing density, the top of the bubble as a function of $\tilde{y}$ is given by
\begin{equation}
z_{u,I}(\tilde{y}) = -2 H \ln (1-\tilde{y}/2)
\end{equation}
and the bottom of the bubble is simply $z_{d,I}(\tilde{y})=-z_{u,I}(\tilde{y})$. In the other case, the coordinate $z_u$ shows an offset of $z_0$
and we derive
\begin{equation}
z_{u,II}(\tilde{y}) = -2 H \ln (1-\tilde{y}/2) + z_0
\end{equation}
and for the bottom of the bubble
\begin{equation}
z_{d,II}(\tilde{y}) = -2 H \ln (1+\tilde{y}/2) + z_0.
\label{ezd}
\end{equation}
Furthermore, the half-width maximum extension of the bubble, $r_{\rm{max}}$, parallel to the galactic disk is found, where 
$\frac{\partial r}{\partial z} = 0$,
\begin{equation}
r_{\rm{max}}(\tilde{y}) = 2H \arcsin \frac{\tilde{y}}{2},
\label{ermax}
\end{equation}
which is valid for both density laws.\\
Obviously, the top of the bubble reaches infinity by the time where $\tilde{y}$ approaches the value 2.
This happens, because the newly swept-up mass asymptotically goes to zero, due to the exponential decrease in density.
The solution then breaks down, because numerically the shock reaches infinity in a finite time. Also the remnant volume goes to infinity (see Figs.~\ref{SBsymm} 
and \ref{SBoff}). In reality, a non-zero
density at infinity leads to restricted values of the shock front velocity \citep{kp98}. 
%
Actually, blow-out occurs earlier in the evolution of a bubble with values of $\tilde{y}$ well below 2.
%
%
%
\section{Results}
We use different energy input schemes to calculate blow-out timescales, the energy input that is needed for blow-out, and the instability timescales.
In the first and simplest case, a single, huge explosion forms the superbubble like in the original version of \citet{k60}. 
A time dependent energy input rate is implemented in the next model, where the number and sequence of SN explosions in a star cluster is given by an initial mass function.
Finally, the wind model uses a constant energy input rate to drive the expansion of the bubble, which is just a special case of the IMF-model. 
In all calculations, a cluster is defined to have at least two member stars.\\
We apply these models to two cases of ISM density distribution. 
First, we study the evolution of a bubble in an ISM representing the Lockman layer of a Milky-Way-type galaxy (number density $n_0 = 0.5 \,$cm$^{-3}$, scale height $H = 500 \,$pc) 
with the center of the explosion placed in the midplane and exponentially, symmetrically decreasing density above and below the disk (Eq.~(\ref{erho1})). 
In the second case, the star cluster is displaced by $z_0$ from the galactic midplane and the SB is expanding into a rather dense ($n_0 = 10 \,$cm$^{-3}$), 
low-scale height ($H = 100 \,$pc) medium described by a simple exponential law (cf. Eq.~(\ref{erho2})).
In this off-plane configuration the bubble shall not expand below the midplane before blow-out (i.e. roughly the time when RTIs start to appear in the shell), 
thus it is not affected by the increasing density below the plane until that time, which results in a one-sided blow-out.
In other words, the value of $z_0$ has to correspond to the absolute value of the 
coordinate $z_d(\tilde{y})$ (Eq.~(\ref{ezd})) in the case of 
an unshifted bubble ($z_0=0$) at the time when the acceleration sets 
in. Values of $z_d(\tilde{y})$ at $y=\tilde{y}_{\rm{acc}}$ range between $0.45\, $H and $1.0 \,$H, thus we take an 
average value and put the explosion at $z_0=0.7 \,$H\footnote{\label{foot:1}The density at the site of the explosion at $z_0=0.7 \,$H 
is already reduced by one half, i.e. $\sim 5 \,$cm$^{-3}$.} in order to be able to compare the models.
This corresponds very well to MM88's criterion for 'one-sided superbubbles': bubbles for which 
the association is found above 0.6~H blow out on one side of the disk only, and the bottom of the bubble should be decelerating more strongly than a spherical one would do.
%

%
%
%
\subsection{Thermal energy}
\citet{bjm99} show that the thermal energy in the hot interior of a SB expanding into an exponentially stratified medium is very close to the value 
in the case of a homogeneous ISM until about four times the dimensionless timescale $t/t_D$, which is rather late in the evolution of the superbubble. 
We find that at the time of fragmentation $\sim 2 -3 \, t/t_D$ are reached (see Table~\ref{frag}). 
Thus, we can estimate the thermal energy in the region of hot shocked gas following the calculations of \citet{w77} for a wind-blown bubble in a uniform ISM.
These take into account the equations of energy and momentum conservation, as well as the radius of a spherical bubble. 
The inner shock, where the energy conversion takes place is always roughly spherical, since it is close to the energy source. 
However, with increasing time, the dynamics of a bubble in an exponentially stratified medium will differ from that in a
homogeneous medium, where no blow-out will happen.
\subsubsection{SN-model} 
All SNe explode at the same time. We find that the fraction of the total SN-energy converted into thermal energy at the inner 
shock is $E_{\rm{th,\, SN}}(t) = 2/3 \cdot E_{\rm{SN}} \cdot N_{\rm{SN}} $ (see~Appendix~\ref{app1}). $N_{\rm{SN}}$ is the total number of SNe, with each explosion releasing 
a constant energy of $E_{\rm{SN}}=10^{51} \,$erg.
\subsubsection{IMF- and wind-model} 
A time-dependent energy input rate $L_{\rm{SB}}(t) = L_{\rm{IMF}} \cdot t^{\, \delta}$ is used, where 
the energy input rate coefficient $L_{\rm{IMF}}$ and the exponent $\delta$ are characterized by the slope of the IMF and the main sequence 
lifetime of massive stars. The calculations of the time-dependent energy input rate follow \citet{bb02} and \citet{f06} and shall be presented here briefly.
The IMF describes the differential number of stars in a mass interval $(m, m+dm)$ by a power law
\begin{equation}
\frac{dN}{dm} = N_0 \cdot m^{\Gamma-1} \, ,
\label{edN}
\end{equation}

\noindent
where $\Gamma$ is the slope of the IMF and $m$ is always given in solar mass units.
Integration from a lower mass limit $m_l$ to an upper mass limit $m_u$ 
gives the cumulative number of stars or -- having $m_l \geq 8 M_{\odot}$ -- the number of OB stars to explode as SNe in this stellar mass range
\begin{equation}
N_{\rm{OB}} = \frac{N_0}{\Gamma} \cdot  \left. m^{\Gamma} \right|_{m_l}^{m_u} \, .
\end{equation}

\noindent
In our general treatment of the IMF either $N_{\rm{0}}$ or $m_u$ will be given and in order to get $N_{\rm{OB}}$ from the equation above, a 
correlation between $N_{\rm{0}}$ and $m_u$ is needed. 
We use integer mass bins and simply 
fix the number of stars in the last mass bin $N(m_u-1, m_u) = N_{\rm{OB}} = 1$, i.e. 
there is exactly one star in the mass bin of the most massive star. Hence the normalization constant is 
$ N_0 = 1 \cdot \Gamma/ \bigl( m_u^{\, \Gamma}- (m_u-1)^{\Gamma} \bigr)$. 
When dealing with a real association, the number of stars $N_{\rm{OB}}$ in a certain mass range ($m_l$, $m_u$) can be deduced from observations and thus the normalization constant will be estimated. Moreover, 
using this information is statistically more relevant than using THE most massive star of the cluster because the distribution of stars in a real clusters may not follow integer mass bins.\\
Since the distribution of the stars by their mass is given by the IMF, and $L_{\rm{SB}}$ follows the time-sequence of massive stars 
exploding as SNe, we get the energy input rate $L_{\rm{SB}} (t) = d/dt \bigl[ \bigl( N(m) \cdot E_{\rm{SN}} \bigr) \bigr] $. 
We just need to express the function $N(m)$ as a time-sequence, thus we treat the number of stars between $(m, m+dm)$ as a function of mass and use 
the main sequence lifetime $(t, t-dt)$ of massive stars

\begin{equation}
L_{\rm{SB}} (t) = E_{\rm{SN}} \frac{dN(m)}{dm} \cdot \left( - \frac{dm}{dt} \right) \, .
\label{eLSB1}
\end{equation}

\noindent
The main sequence lifetime and the mass of a star are connected through
$
t(m) = \kappa \cdot m^{-\alpha} 
$
or
$ m(t) = (t/\kappa)^{-1/\alpha}.$
The values of \citet{f06} are used throughout this paper, where $\kappa=1.6\cdot 10^8 \,$yr and $\alpha=0.932$.  
The time-derivative of $m(t)$ together with $dN/dm$ from Eq.~(\ref{edN}) are inserted into Eq.~(\ref{eLSB1})

\begin{equation}
L_{\rm{SB}}(t) = \frac{E_{\rm{SN}} \cdot N_0}{\alpha \cdot \kappa} \left( \frac{t}{\kappa} \right)^{-(\Gamma/\alpha+1)} \, .
\label{eLSB2}
\end{equation}

\noindent
Since $L_{\rm{SB}} (t) \propto t^{\, \delta}$, the exponent must be $\delta = - \left( \Gamma/\alpha + 1 \right)$.
After summarizing the constants one obtains the energy input rate coefficient
\begin{equation}
L_{\rm{IMF}} = \frac{E_{\rm{SN}} \cdot N_0 \cdot (\kappa)^{\Gamma/\alpha}}{\alpha}.
\label{eLIMF}
\end{equation}
%
The full equation of the thermal energy as a function of time for the IMF-model is (see Appendix~\ref{app1} for details)

\begin{equation}
E_{\rm{th,\, IMF}}(t) = \frac{5}{7 \delta +11} \cdot L_{\rm{IMF}} \cdot t^{{\delta}+1}.
\label{eEthimf}
\end{equation}

\noindent
Using an IMF yields a more realistic framework for galactic SB expansion compared to a simplified point explosion.
Therefore, we
investigate the effect of changing the slope of the IMF in our calculations. 
We compare three different IMF-slopes for massive stars:
$\Gamma_1=-1.15$ \citep{bg03}, $\Gamma_2=-1.35$ \citep{s55}, and $\Gamma_3=-1.7$ \citep{b94,sca86} resulting in $\delta_1=0.23$, $\delta_2=0.45$, and $\delta_3=0.82$.\\
With an exponent $\delta=0$, Eq.~(\ref{eEthimf}) respresents the thermal energy in case of a constant energy input rate,
where the number of all SNe is averaged over the whole lifetime of the association, thus this model simply has a slope of $\Gamma_0=-0.932$ (wind-model). 
Now the larger time intervals between SN-explosions of stars with higher main sequence lifetime compensate the growing number of SNe per mass interval
going to lower mass stars as it would be the case in the IMF-model.
With a constant energy input rate $L_w$, our result (Eq.~(\ref{eEthimf})) checks with Weaver et al.'s relation of  
$E_{\rm{th,w}}(t) = 5/11 \cdot L_w \cdot t$.
%
%
%
%
\subsection{Evolution of a superbubble until blow-out}
The transformed time variable (Eq.~(\ref{ey})) contains not only the energy input and the ambient density, but also the volume of a bubble at a 
given time. Since this equation gives us an explicit correlation between
time $t$ and the time variable $\tilde{y}$ it is possible to describe the volume of the bubble as a function of $\tilde{y}$
instead of being dependent on time.
In order to proceed with our analytical description we need to find simple expressions of the volume for the two cases of density distribution.
\paragraph{Symmetric model} The bubble contour on the $+z$-side of the midplane can be approximated by being part of an ellipse with 
a semimajor axis of 
\begin{equation}
a(\tilde{y}) = \frac{z_{u,I}(\tilde{y}) -  z_b(\tilde{y})}{2}
\end{equation}
and $z_b(\tilde{y}) =-2 H \ln (1+\tilde{y}/2)$ corresponding to the bottom of an unshifted bubble in a pure exponential atmosphere.
The semiminor axis $b(\tilde{y}) $ is equal to $r_{\rm{max}}(\tilde{y})$ (see Eq.~(\ref{ermax})).
The center of the ellipse is located on the $z$-axis
\begin{equation}
z_r(\tilde{y}) = z_{u,I}(\tilde{y}) - a(\tilde{y}) \, ,
\end{equation}
such that the ellipse is given by 
\begin{equation}
r_{\rm{ell}}(\tilde{y},z)=  \sqrt{b^2(\tilde{y}) -  b^2(\tilde{y}) / a^2(\tilde{y}) \cdot (z-z_r(\tilde{y}))^2} \, .
\end{equation}

\noindent
Rotating this curve around the $z$-axis and multiplying by two results in the total volume of the superbubble
\begin{equation}
V_I(\tilde{y})=  2 \cdot \pi \cdot \int_0^{z_{u,I}} r^2_{\rm{ell}}(\tilde{y},z) \, dz \, .
\label{eVI}
\end{equation}

\noindent
The approximation works very well and we find a deviation of Eq.~(\ref{eVI}) from the numerical integration of the volume according to Eq.~(\ref{eomega}) of only $\sim$ 2.2 $\%$
at very late stages of evolution ($\tilde{y} = 1.9$).
%
%
\begin{figure*}
\centering
\includegraphics[angle=270,width=\textwidth]{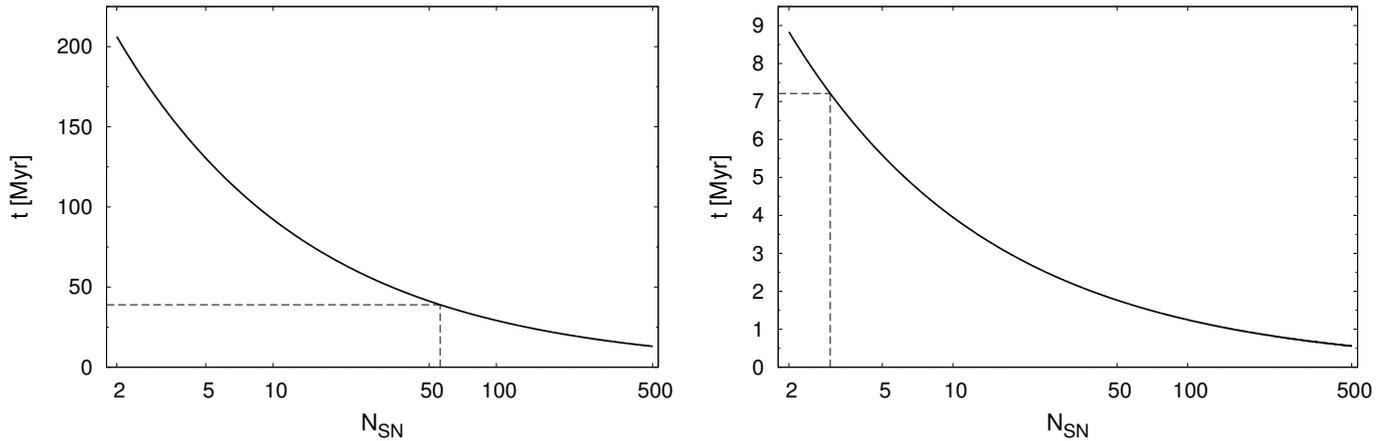}
\caption{\label{time_SN} Blow-out timescales at the coordinate $z_u$ for the SN-model. Left: double exponential layer with 
$H = 500 \, $pc and $n_0=0.5\,$cm$^{-3}$ (symmetric model); 
right: exponentially stratified medium with $H = 100 \, $pc and $n_0=10\,$cm$^{-3}$ (off-plane model). The dashed lines indicate 
the number of SNe required for blow-out and the corresponding blow-out time.}
\end{figure*}
%
%
%
%
%

\paragraph{Off-plane model} The volume will be approximated by treating the 3D shape of the bubble as a prolate ellipsoid \citep[see also][]{mc99}. 
It is not dependent on the offset $z_0$, thus  
the bubble volume as a function of $\tilde{y}$ reads
\begin{equation}
V_{II}(\tilde{y}) = \frac{4 \pi}{3} \, a(\tilde{y}) \, b^2(\tilde{y}) = \frac{16 \pi H^3}{3} \cdot \arcsin^2 \left(\frac{\tilde{y}}{2} \right) \cdot
 \ln\frac{1 + \frac{\tilde{y}}{2}}{1-\frac{\tilde{y}}{2}} \, .
\label{eVII}
\end{equation}

\noindent
At $\tilde{y} = 1.9$, the analytical result differs from the numerical value according to Eq.~(\ref{eomega}) by only $\sim 1.8\, \%$.\\
\\
Obviously, the parameter $\tilde{y}$ is important for estimating the evolutionary status of a bubble, since it relates the size of the bubble 
to time.
We can now calculate the age $t$ of a bubble at any value of $\tilde{y}$.
Replacing $\frac{dy}{dt}= H \cdot \frac{d\tilde{y}}{dt}$ and using the non-dimensional volume $\tilde{V}_{\rm{I,II}}(\tilde{y}) = V_{\rm{I,II}}(\tilde{y})/H^3$ 
instead of $\Omega \, (t)$ helps to obtain a useful relation of the time derivative of Eq.~(\ref{ey})
\begin{equation}
\frac{d\tilde{y}}{dt} = \frac{1}{H^{5/2}}\cdot\sqrt{\frac{\gamma^2 - 1}{2}\cdot \frac{E_{\rm{th}}}{\rho_{0,1} \cdot \tilde{V}_{\rm{I,II}}(\tilde{y})}} \, .
\label{edydt}
\end{equation}
When performing the integration, the thermal energy for each model, i.e.$\,$for different kinds of energy input,
has to be replaced by the corresponding formula. Substituting $\beta = \sqrt{(\gamma^2 - 1)/2}$
and integrating yields the general expression
\begin{equation}
\int_0^t \sqrt{E_{\rm{th}}} \, dt'= \sqrt{\rho_{0,1}} \cdot \frac{ H^{5/2}}{\beta} \cdot \int_0^{\tilde{y}} \sqrt{\tilde{V}_{\rm{I,II}}(\tilde{y}')} \, d\tilde{y}' \, ,
\label{eintE}
\end{equation}
which will be solved separately for all three models in the remainder of this subsection.
In order to simplify the integration on the right hand side, and especially to avoid a double integral in the case of $V_I(\tilde{y})$ in the calculations to follow, we use a series expansion of 
the bubble's volume:
\begin{equation}
\int_0^{\tilde{y}} \sqrt{\tilde{V}_{\rm{I}}(\tilde{y}')} \, d\tilde{y}' = 0.8187 y^{5/2} + 0.1096 y^{7/2} + 0.0299 y^{9/2}   \, ,
\end{equation}
\begin{equation}
\int_0^{\tilde{y}} \sqrt{\tilde{V}_{\rm{II}}(\tilde{y}')} \, d\tilde{y}' = 0.8187 y^{5/2} + 0.0379 y^{9/2} + 0.0037 y^{13/2} \, ,
\end{equation}
In both cases of density distribution we expand $\sqrt{\tilde{V}_{\rm{I,II}}(\tilde{y}')}$ until 43rd order to make sure the simplified integral has a deviation of $\ll 1\, \%$ from the numerical value 
at a time of $\tilde{y} = 1.9$.\\ 
For further investigation of the blow-out phenomenon we need to know the velocity and acceleration 
of the shock front. It is possible to derive these properties analytically at certain points of the bubble's surface, where explicit equations exist. 
The velocity can be calculated for top, bottom and maximum radial extension of the bubble, and also at $r(\tilde{y},0)$ for the symmetric model as well as at 
$r(\tilde{y},z_0)$ for the off-plane model. We will concentrate on calculating the velocity at the top of the SB (which is equal to the absolute value at the bottom for the symmetric model) in this paper, since this is crucial in determining if 
blow-out happens or not. 
For all models, the velocity at the top of the bubble is given by 
\begin{equation}
\dot{z}_u(\tilde{y})=\frac{dz_u}{dt} = \frac{dz_u}{d\tilde{y}} \frac{d\tilde{y}}{dt}
\label{edotzu}
\end{equation}
with the derivative of $z_u$ with respect to $\tilde{y}$ 
\begin{equation}
\frac{dz_u}{d\tilde{y}} = \frac{d}{d\tilde{y}} \Bigl( -2H \cdot \ln(1 -\tilde{y}/2) \Bigr) =  \frac{H}{1 - \tilde{y}/2} \, .
\label{edzudy}
\end{equation}
Whereas the velocity given by Eq.~(\ref{evn}) depends on time $t$ and the coordinate $z$, the velocity in Eq.~(\ref{edzudy}) is only dependent on $\tilde{y}$. This makes it easier to find general 
results in terms of the $y$-parameter at the time of blow-out for each model specification.
The second derivative of $z_u$ gives us the acceleration at this coordinate
\begin{equation}
\ddot{z}_u(\tilde{y}) = \frac{d\dot{z}_u(\tilde{y})}{dt}=\frac{d\dot{z}_u}{d\tilde{y}}\frac{d\tilde{y}}{dt} \, .
\label{eddotzu}
\end{equation}

\begin{figure*}
\centering
\includegraphics[angle=270,width=\textwidth]{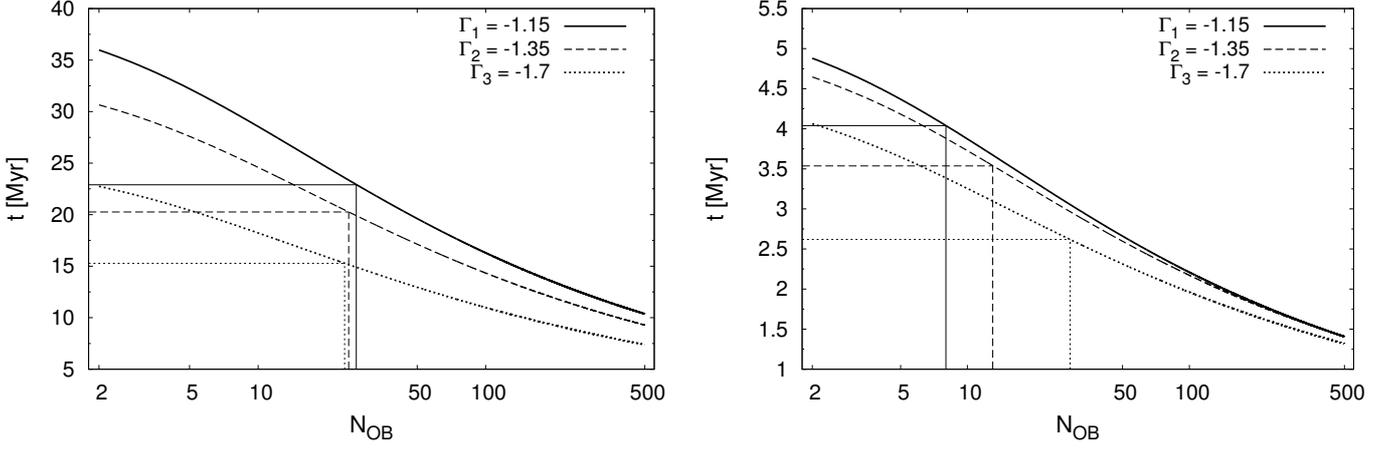}
\caption{\label{time_IMF} 
Same as Fig.~\ref{time_SN}, but for the IMF-model (left: symmetric model; right: off-plane model). Thin lines indicate the number of OB-stars needed for blow-out. $\Gamma$ is the IMF exponent (for details, see text).}
\end{figure*}

\begin{figure*}
\centering
\includegraphics[angle=270,width=\textwidth]{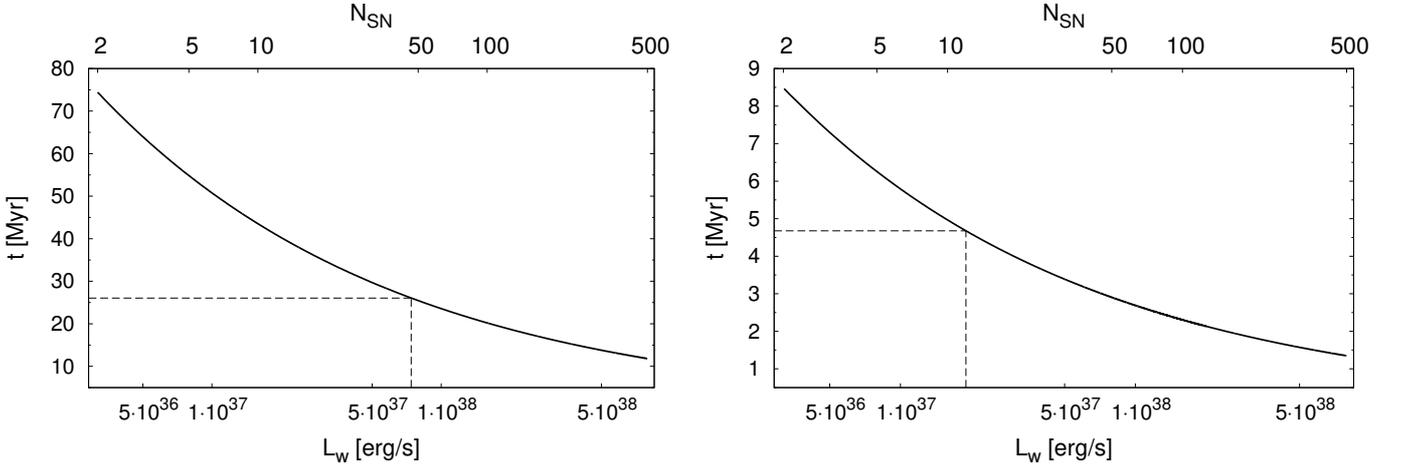}
\caption{\label{time_wind} Same as Fig.~\ref{time_SN}, but as a function of a constant energy input rate (left: symmetric model; right: off-plane model).
The number of SNe is obtained by converting the wind luminosity to an energy input averaged over a time interval of 20 Myr, from the explosion of the most massive star until the last SN.}
\end{figure*}

\noindent
The calculation of $d\tilde{y}/dt$ (Eq.~(\ref{edydt})) and thus $d\dot{z}_u/d\tilde{y}$ need to be done separately for each model, which is 
shown in Appendix~\ref{app2}.\\
Once velocity and acceleration are obtained, we get the value of $\tilde{y}_{\rm{acc}}$, where the velocity of the top of the bubble has 
its minimum, i.e. the transition from deceleration to acceleration along the density gradient. 
The value of the dimensionless variable $\tilde{y}_{\rm{acc}}$ is the same for all SBs of each model, but corresponds to a different time in the evolution of a bubble 
depending on the ISM parameters and the energy input.
The time interval until $\tilde{y}_{\rm{acc}}$ is called blow-out timescale (see Figs.~\ref{time_SN}, \ref{time_IMF}, and \ref{time_wind} for SN-, IMF-, and wind-model,
respectively).\\
By fixing the velocity of the outer shock at the time $\tilde{y}_{\rm{acc}}$, the energy input required for 
blow-out of the disk can be derived. We assume that for $M \geq 3$ at $z_u$ the shock is sufficiently strong and the blow-out condition is fulfilled.
Thus, the velocity of the shock has to be at least $3 \cdot c_s$ with respect to an ambient ISM at rest.
Using a temperature of the surrounding medium of $T \sim 6000\,$K typical for the Lockman layer of the Galaxy \citep{clp02}, the minimum 
velocity corresponds to $v_{\rm{acc}} = 3 \cdot \sqrt{k_B T/\bar{m}} \cong 2 \cdot 10^6 \,$cm/s, where $k_B$ is Boltzmann's constant; the mean atomic mass is $\bar{m}=\mu m_H$ in a gas with mass density $\rho_0=n_0 \cdot \bar{m}$ 
where the mean molecular weight of the neutral
ISM of $\mu=1.3$ and the hydrogen mass of $m_H=1.7\cdot 10^{-24} \,$g are used. 
This criterion is valid for the low-scale height, high-density ISM as well, because one-sided SBs blow out of the dense disk and start to accelerate into the halo at $\sim 2 \, $H $= 
200 \,$pc (Table \ref{accel}). 
At these distances from the midplane the presence of the warm gas layer already influences the evolution of the bubble.
The expressions for the minimum energy input are derived in the following subsections.
%
%
%
%
%
%
%
%
%
%
%
\subsubsection{SN-model}
Since the evolution of the bubble depends on the scale height $H$, on the density $\rho_{0,1}$ of the ambient medium near the 
energy source, and on the energy $E_{\rm{th,SN}}$, a timescale (in units of seconds) can be constructed from these quantities
\begin{equation}
t_{\rm{SN}} = \sqrt{\frac{\rho_{0,1} \cdot H^5}{E_{\rm{th,\, SN}}}} \, .
\label{tSN}
\end{equation}

\noindent
Solving Eq.~(\ref{eintE}) yields the time $t$ elapsed since the explosion
for any chosen value of $\tilde{y}$. For a constant value of energy released by a number of supernovae, the integration on the left hand side is simple.
%
%
Expressing the time $t$ as a function of $\tilde{y}$, rearranging the equation and making use of the timescale (Eq.~(\ref{tSN})) yields the following equation
\begin{equation}
t(\tilde{y}) = \frac{t_{\rm{SN}}}{\beta} \cdot \int_0^{\tilde{y}} \sqrt{\tilde{V}_{\rm{I,II}}(\tilde{y}')} \, d\tilde{y}' \, .
\end{equation}

\noindent
For the calculation of the velocity, we first express Eq.~(\ref{edydt}) in terms of the timescale $t_{\rm{SN}}$ 
\begin{equation}
\frac{d\tilde{y}}{dt} =  \sqrt{\frac{E_{\rm{th,\, SN}}}{H^5 \cdot \rho_{0,1}}} \cdot \frac{\beta}{\sqrt{\tilde{V}_{\rm{I,II}}(\tilde{y})}} =
\frac{1}{t_{\rm{SN}}} \cdot  \frac{\beta}{\sqrt{\tilde{V}_{\rm{I,II}}(\tilde{y})}} \, .
\label{edydtSN}
\end{equation}

\noindent
Now, using the expressions from above, the velocity at the top of the bubble can be written as
\begin{equation}
\dot{z}_u(\tilde{y}) = \frac{\beta}{t_{\rm{SN}}} \cdot \frac{H}{1 - \tilde{y}/2} \cdot \frac{1}{\sqrt{\tilde{V}_{\rm{I,II}}(\tilde{y})}} \, .
\label{edzuSN}
\end{equation}

\noindent
The calculation of the acceleration at the top of the bubble is found in App.~\ref{app2}.
The critical value $\tilde{y}_{\rm{acc}}$, where the acceleration of the outer shock starts, the corresponding dimensionless timescale and the coordinate of the top of 
the bubble at this time are summarized in Table \ref{accel} for both cases of density distribution. 
%
Fig.~\ref{time_SN} shows the age of the bubble at the transition from deceleration to acceleration as a function of the number of SN-explosions in the range of $2-500$ SNe. 
The dashed lines indicate the number of SNe that are 
required for blow-out for each kind of density distribution and the corresponding timescale for blow-out. 
In the case of the symmetric density law with $H = 500 \, $pc and $n_0=0.5\,$cm$^{-3}$ about 56 SNe have to explode 
at once and the acceleration starts $\sim$ 40 Myr after the initial explosion. Only three SNe are sufficient for an off-plane explosion in a 
pure exponential atmosphere at $z_0=0.7\,$H with $H = 100 \, $pc and $n_0=10\,$cm$^{-3}$. In this case, blow-out happens after $\sim$ 7.2 Myr.\\
In order to obtain the general dependence of the minimum number of SNe on the properties of the surrounding medium at the coordinate $z_u$, the thermal energy $E_{\rm{th,\, SN}}$
appearing in $t_{\rm{SN}}$ (Eq.~(\ref{tSN})) is replaced by $2/3 \cdot E_{\rm{SN}} \cdot N_{\rm{SN}}$ as derived in Appendix~\ref{app1} and the velocity at $z_u$ (Eq.~(\ref{edzuSN})) is solved for $N_{\rm{SN}}$: 
\begin{equation}
N_{\rm{SN,\, blow}}(H, n_0)=  \frac{n_0\cdot k \cdot \bar{m} \cdot H^3 \cdot \tilde{V}_{\rm{I,II}}(\tilde{y})}{2 \cdot E_{\rm{SN}}/3}  
\left( \frac{v_{\rm{acc}} \cdot (1-\frac{\tilde{y}}{2})}{\beta}
\right)^2.
\label{eNSN}
\end{equation}

\noindent
The velocity is fixed to be $v_{\rm{acc}}= \dot{z_u}(\tilde{y}=\tilde{y}_{\rm{acc}}) \cong 2 \cdot 10^6 \,$cm/s at the time of blowout and the constants  
$k=1$ for the symmetric model and $k= \exp(-z_0/H)$ for the off-plane model, respectively, are used.
%
%
%
%
%
%
%
%
%
%
%

%
%
%
%
%
%
%
%
\subsubsection{IMF- and wind-model}
Next, we want to calculate the age, velocity and acceleration of a bubble driven by a time-dependent energy input rate, where different slopes of the IMF will be used.
As a special case of this model, we can describe the evolution of a bubble powered by a constant energy input rate. 
Using a procedure similar as above, the characteristic timescale for the IMF-model in terms of the energy input rate coefficient 
$L_{\rm{IMF}}$, the density and the scale height of the interstellar medium is found to be 
\begin{equation}
t_{\rm{IMF}} = \left ( \frac{\rho_{0,1} \cdot H^5}{L_{\rm{IMF}}} \right )^{1/(\delta + 3)}.
\label{etIMF}
\end{equation}

\noindent
After inserting the thermal energy derived for the IMF-model (Eq.~(\ref{eEthimf})) into Eq.~(\ref{eintE}) and solving the integral,
one gets the time as a function of $\tilde{y}$
\begin{equation}
t(\tilde{y}) = t_{\rm{IMF}} \cdot \left( \frac{(\delta + 3)^2 (7 \delta + 11)}{20 \, \beta^2} \right) ^{\frac{1}{\delta + 3}}
\cdot \left( \int_0^{\tilde{y}} \sqrt{\tilde{V}_{\rm{I,II}}(\tilde{y}')}\,d\tilde{y}' \right)^{\frac{2}{\delta + 3}}.
\label{etyimf}
\end{equation}
\begin{table*}
\centering
\begin{tabular}{cccccccccc}
\hline
\hline\\[-0.9em]
 & \multicolumn{3}{c}{N$_0$(N$_{\rm{OB}}$)}    & \multicolumn{3}{c}{N$_{\rm{OB}}$(N$_0$)} &  \multicolumn{3}{c}{m$_u$(N$_{\rm{OB}}$)} \\[0.3em]
\multicolumn{1}{c}{$\Gamma$}  &\multicolumn{1}{c}{$a_1$}   & \multicolumn{1}{c}{$b_1$}  & \multicolumn{1}{c}{$c_1$} &   \multicolumn{1}{c}{$a_2$}  
& \multicolumn{1}{c}{$b_2$} &  \multicolumn{1}{c}{$c_2$}  & \multicolumn{1}{c}{$a_3$}  & \multicolumn{1}{c}{$b_3$} &   \multicolumn{1}{c}{$c_3$}    \\
\hline\\[-0.8em]
-1.15 &  20.395  &  0.934    &  83.700   &   0.037  &  1.076    &  -4.720  &  2.549  & 0.495   &  6.451     \\ 
-1.35 &  31.561  &  0.955    &  134.738  &   0.025  &  1.055    &  -4.566  &  2.719  & 0.464   &  6.324     \\
-1.7  &  76.445  &  0.963    &  265.063  &   0.011  &  1.041    &  -3.772  &  2.938  & 0.421   &  6.181     \\
\hline                                                                    
\end{tabular}
\caption{\label{N0NOB} 
Polynomial fits of three typical IMF-relations.
Different IMF slopes are given in the first column. The coefficients of each polynomial fit appear in the remaining columns.
The first relation $N_0(N_{\rm{OB}}) = a_1 \cdot N_{\rm{OB}}^{b_1} + c_1$ is used to give blow-out timescales as a function of the number of OB-stars (Fig.~\ref{time_IMF}), 
with the help of the second relation $N_{\rm{OB}}(N_0)=a_2 \cdot N_0^{b_2}+ c_2$, Eq.~(\ref{eN0min}) can be better understood, and finally
$m_u(N_{\rm{OB}})=a_3 \cdot N_{\rm{OB}}^{b_3} + c_3$ is used for the fragmentation timescales (Fig.~\ref{frag_imf}).
} 
\end{table*} 

\noindent
Moreover, inserting the thermal energy for this model given by Eq.~(\ref{eEthimf}) into Eq.~(\ref{edydt}) and using the time $t$ and the timescale $t_{\rm{IMF}}$ gives
\begin{equation}
\begin{split}
\frac{d\tilde{y}}{dt} &= \sqrt{\frac{E_{\rm{th,IMF}}(t)}{H^5 \cdot \rho_{0,1}}} \cdot \frac{\beta}{\sqrt{\tilde{V}_{\rm{I,II}}(\tilde{y})}} =\\
&= \sqrt{\frac{\left(\frac{5}{7 \delta + 11}\right) \cdot L_{\rm{IMF}} \cdot t^{\delta+1}}{H^5 \cdot \rho_{0,1}}} \cdot \frac{\beta}{\sqrt{\tilde{V}_{\rm{I,II}}(\tilde{y})}}=\\
&= \frac{\beta}{t_{\rm{IMF}}} \cdot \sqrt{ \frac{5}{7 \delta + 11}} 
\cdot \left( \frac{(\delta + 3)^2 \cdot (7 \delta + 11)}{20 \, \beta^2} \right)^{\frac{(\delta + 1)\cdot \epsilon}{2}} \cdot \\
&\cdot
\left( \int_0^{\tilde{y}} \sqrt{\tilde{V}_{\rm{I,II}}(\tilde{y}')}\, d\tilde{y}'\right)^{(\delta + 1)\cdot \epsilon}
\frac{1}{\sqrt{\tilde{V}_{\rm{I,II}}(\tilde{y})}} \, ,
\end{split}
\label{edydtimf}
\end{equation}

\noindent
where $\epsilon= 1/(\delta + 3)$.
Using further substitutions
\begin{equation}
\xi = \beta \cdot \sqrt{\frac{5}{7 \delta + 11}} \cdot \left(\frac{(\delta + 3)^2 \cdot (7 \delta + 11)}{20 \, \beta^2}\right)^{\frac{(\delta + 1)\cdot \epsilon}{2}}
\end{equation}

\noindent
and
\begin{equation}
\bar{\delta} = (\delta + 1)\cdot \epsilon
\end{equation}

\noindent
yields a simplified expression of Eq.~(\ref{edydtimf})
\begin{equation}
\frac{d\tilde{y}}{dt} = \frac{\xi}{t_{\rm{IMF}}} \cdot \left(\int_0^{\tilde{y}} \sqrt{\tilde{V}_{\rm{I,II}}(\tilde{y}')}\, d\tilde{y}' 
\right)^{\bar{\delta}} \frac{1}{\sqrt{\tilde{V}_{\rm{I,II}}(\tilde{y})}} \, .
\label{edydtimf2}
\end{equation}

\noindent
As it was done in the previous model, multiplying this equation by $dz_u/d\tilde{y}$ (Eq.~(\ref{edzudy})) results in the velocity at the top of the 
expanding superbubble 
\begin{equation}
\dot{z}_u(\tilde{y}) = \frac{\xi}{t_{\rm{IMF}}} \cdot \frac{H}{1 - \tilde{y}/2} \cdot \left(\int_0^{\tilde{y}} 
\sqrt{\tilde{V}_{\rm{I,II}}(\tilde{y}')}\,d\tilde{y}' \right)^{\bar{\delta}}
\cdot \frac{1}{\sqrt{\tilde{V}_{\rm{I,II}}(\tilde{y})}} \, .
\label{edzudtimf}
\end{equation}

\noindent
For the calculation of the acceleration $\ddot{z}_u(\tilde{y})$, see App.~\ref{app2}.
Now that general expressions for velocity and acceleration of the bubble are known, we can estimate when a bubble starts to accelerate 
into the halo. The results for the IMF-model using different slopes and for the wind-model are listed in Table \ref{accel}.\\
In order to present the blow-out timescales for superbubbles driven by a time-dependent energy input rate as a function of the number of OB-stars,  $N_{\rm{OB}}$, 
instead of the normalization constant $N_0$, which is contained in $t_{\rm{IMF}}$ (Eq.~(\ref{etyimf})), we have to 
make use of the fit $N_0(N_{\rm{OB}})$ from Table \ref{N0NOB}. The coefficients of the power-law function $N_0(N_{\rm{OB}}) = a_1 \cdot N_{\rm{OB}}^{b_1} + c_1$
found with a nonlinear least-squares 
fit provide an excellent fit to the relation with errors less than $1\,\%$ for $2 \le N_{\rm{OB}} \le 500$
for all IMF-models. The timescales in Fig.~\ref{time_IMF} are shown for $2-500$ association members, where the dashed lines represent the minimum number of 
OB-stars and the corresponding time until blow-out. Using IMFs with $\Gamma_1=-1.15$, $\Gamma_2=-1.35$, and $\Gamma_3=-1.7$, respectively, 
27, 25, and 24 OB-stars are needed with an upper mass limit of 19, 18 and 17 $M_{\odot}$.
The resulting blow-out timescales are 22.8, 20.1, and 15.2 Myr for a symmetric density distribution with $H=500 \,$pc 
and $n_0=0.5 \,$cm$^{-3}$ (Table \ref{accel} $\&$ Fig.~\ref{time_IMF}, left).
In the case of an explosion at 70 pc above the plane in an ISM with $H=100 \,$pc and 
$n_0=10 \,$cm$^{-3}$ (Table \ref{accel} $\&$ Fig.~\ref{time_IMF}, right), the association needs to have at least $\sim$ 8, 13, and 29 massive stars or an upper mass limit of 13, 15 and 18 $M_{\odot}$
(same order of IMF-slopes). Approximately 4.0, 3.5, and 2.6 Myr pass from the first SN-explosion until blow-out.
We are interested again in obtaining an analytical expression for the minimum number of OB-stars to get blow-out. Eq.~(\ref{edzudtimf}) needs to be solved for
the normalization constant, after $L_{\rm{IMF}}$ was replaced by Eq.~(\ref{eLIMF}), which yields  
\begin{equation}
N_{\rm{0,\, blow}}(H, n_0) = \frac{n_0 \cdot k \cdot \bar{m} \cdot H^{2-\delta} \cdot \alpha} {E_{\rm{SN}} \cdot \kappa^{\Gamma/\alpha}} 
\cdot \left( \frac{ v_{\rm{acc}}\cdot \sqrt{\tilde{V}_{\rm{I,II}}(\tilde{y})} \cdot (1 - \frac{\tilde{y}}{2} ) }{\xi \cdot \left( \int_0^{\tilde{y}} 
\sqrt{\tilde{V}(\tilde{y}')} \, d\tilde{y}' \right)^{\bar{\delta}} }   \right)^{\delta +3}  
\label{eN0min}
\end{equation}

\noindent
for any set of the parameters scale height and ISM-density.
This has to be converted to a number of stars $N_{\rm{OB,\, blow}}$ by using the fit $N_{\rm{OB}}(N_0)=a_2 \cdot N_0^{b_2}+ c_2$ presented in Table \ref{N0NOB}, which is obtained with the 
same fitting procedure as before. We find that our fits are very good for all IMF-slopes with an average deviation of $\sim 1.4\,\%$ in the range of $2-500$ OB-stars.
In Fig.~\ref{vergl} we compare the number of stars required for blow-out as a function of the scale height using certain values of the density  
for the SN- and the IMF-models. The discussion of the results is found in the next section.\\
Using a constant energy input rate ($\delta=0$), we obtain a dimensionless wind coefficient
\begin{equation}
L_{\rm{0, \, blow}}(H, n_0)= \frac{n_0 \cdot k \cdot m \cdot H^{2} \cdot \alpha \cdot \kappa} {E_{\rm{SN}}} 
\cdot \frac{ \left( v_{\rm{acc}}\cdot \sqrt{\tilde{V}_{\rm{I,II}}(\tilde{y})} \cdot (1 - \frac{\tilde{y}}{2}) \right)^{3}  }{\xi^3 \cdot \int_0^{\tilde{y}} 
\sqrt{\tilde{V}(\tilde{y}')} \, d\tilde{y}' }   
\end{equation}

\noindent
instead of the normalization constant. Following Eq.~(\ref{eLSB2}), we can derive the minimum wind luminosity that fulfills the blow-out-criterion
\begin{equation}
L_{\rm{w,\, blow}}(H, n_0)= \frac{E_{\rm{SN}} \cdot L_{\rm{0,\, min}}(H, n_0)}{\alpha \cdot \kappa}
\label{eLwmin}
\end{equation}

\noindent
in units of erg/s. 
We assume that the wind luminosity is the energy input of all SNe averaged over the period of $\Delta t \approx 20 \,$Myr, the lifetime of 
the association, which is in general the time between the first and the last SN-explosion\footnote{\label{foot:2} Except for a very small 
cluster with member stars of approximately the same mass. In such a case the time span between the SN explosions can be much smaller than 20 Myr.}.
Hence, the minimum wind luminosity can be converted into a total number of stars $N_{\rm{w,blow}}=L_{\rm{w,blow}} \cdot \Delta t/ E_{\rm{SN}}$. 
Fig.~\ref{time_wind} shows the time $t(\tilde{y})$ elapsed since  
$\tilde{y}=0$ until the point of acceleration, for wind luminosities between $3\cdot10^{36}$ and $8 \cdot 10^{38}$ erg/s, corresponding to $\sim 2 - 500 $ SNe 
calculated over a timespan of 20 Myr. 
The values for the minimum wind luminosity and 
corresponding ages of the SB are given by the dashed lines.
For a symmetric expansion into the Lockman layer, an energy input rate of 
$\sim 7.4 \cdot 10^{37} \,$erg/s is necessary for blow-out which corresponds to $\sim$ 47 SNe. Such a bubble needs 26.1 Myr until acceleration begins.
Expansion of a SB into the low-scale height, high-density medium produced by an off-plane explosion at $z_0=0.7\,$H requires $\sim 1.9 \cdot 10^{37} \,$erg/s or 
about 12 SNe and takes $\sim$ 4.7 Myr. 
%
%
%
%
%
%
%
%
%
\subsection{Rayleigh-Taylor instabilities in the shell}
Infinitesimal perturbations at the interface between a denser fluid supported by a lighter fluid in a gravitational field generate waves
with amplitudes growing exponentially with time in the initial phase. For an incompressible, inviscid, non-magnetic fluid, the time
scale $\tau_{\rm{rti}}$, characterizing the growth of the instability, results from a linear stability analysis combined with the conservation 
equations 
\begin{equation}
\tau_{\rm{rti}} = \sqrt{\frac{\lambda}{2 \pi g} \cdot \frac{\rho_2 + \rho_1}{\rho_2 - \rho_1}} \, ,
\end{equation}

\noindent
where $\lambda$ is the perturbation wavelength, $g$ is the gravitational acceleration and $\rho_1$ and $\rho_2$ are the densities of the light and heavy fluid, respectively.
In the case of a superbubble, where the dense shell is accelerated by the hot, tenuous gas, the most important perturbation wavelengths out of the Fourier spectrum are the ones, which are comparable to the thickness of the 
shell $\Delta d(\tilde{y})$ (which is itself a function of time), because these distort the shell so strongly that break-up can occur. Furthermore, identifying the gravitational acceleration with the acceleration at the top of the bubble 
$\ddot{z}_u(\tilde{y})$ results in an instability timescale at this coordinate
\begin{equation}
\tau_{\rm{rti},z_u}(\tilde{y})= \sqrt{ \frac{\Delta d(\tilde{y})}{2 \pi \ddot{z}_u(\tilde{y})} \cdot \frac{\rho_{\rm{sh}}(\tilde{y}) + 
\rho_{\rm{in}}(\tilde{y})} {\rho_{\rm{sh}}(\tilde{y}) - \rho_{\rm{in}} (\tilde{y})} } \, .
\label{etrti}
\end{equation}
%
The density in the shell at the coordinate $z_u$ is always given by $\rho_{\rm{sh}}(\tilde{y}) = 4 \cdot\rho_{0,1} \exp[-z_u(\tilde{y})/H]$ for an adiabatic strong 
shock\footnote{\label{foot:3} This only represents a simplified picture, because the streamlines in an ellipsoidal bubble are not radial, and therefore not parallel to the shock normal. 
But the error is small, because the shell is still thin until break-up.}.
The density of the bubble interior as a function of time is $\rho_{\rm{in}}(\tilde{y})= M_{\rm{ej}}(\tilde{y})/V_{\rm{in}}(\tilde{y})$ with 
$M_{\rm{ej}}(\tilde{y})$ being the ejecta mass and $V_{\rm{in}}(\tilde{y})$ being the volume of the bubble confined by the inner boundary of the shell.
The new semimajor and semiminor axes simply reduced by the thickness of the shell are 
$a_{\rm{in}}(\tilde{y}) = a(\tilde{y}) - \Delta d(\tilde{y})$ and  
$b_{\rm{in}}(\tilde{y}) = b(\tilde{y}) - \Delta d(\tilde{y})$.
To calculate the thickness of the shell we use the fact, that the mass of the gas in a volume $V(\tilde{y})$ of undisturbed ISM is the same as that in the shell of swept-up ISM (if we neglect effects of evaporation, heat conduction or mass loading).
In a symmetrically decreasing density distribution the bubble contour has an hourglass-shape, thus, due to symmetry,
we need to include half of the bubble only in the calculation 
\begin{equation}
\Delta d_{\rm{I}}(\tilde{y}) = \frac{M_{\rm{frac}}(\tilde{y})}  {\int_{0}^{z_u} \rho_{\rm{sh,\rm{I}}}(\tilde{y}) \, dA_{\rm{ell}} (\tilde{y}) } \, .
\label{edI}
\end{equation}

\noindent
Half of the shell mass is $M_{\rm{frac}}(\tilde{y})= \int_{0}^{z_u} \rho_{\rm{I}}(z) \, dV_{I} $, which we obtain by integration over the density gradient.
The bubble's volume, confined by the surface $A_{\rm{ell}}$ of an 
ellipsoid, is included between the coordinates $z=0$ and $z_u$.
Using the new semimajor and semiminor axes -- each itself a function of $\tilde{y}$ -- in the equation of the ellipse means solving a double 
integral in the calculation 
of the volume (Eq.~(\ref{eVI})). However, we can approximate by using an average thickness of the shell over the time in this case. The new 
semimajor axis is as large as $\sim$ 91$ \, \%$ of the regular value and the new semiminor axis $\sim$ 90$ \, \%$ (averaged over the time span $0.1 \leq \tilde{y} \leq 1.9$).\\
In the case of an off-plane explosion the thickness of the shell is given by
\begin{equation}
\Delta d_{\rm{II}}(\tilde{y}) = \frac{M_{\rm{sh}}(\tilde{y})}  {\int_{z_d}^{z_u} \rho_{\rm{sh,\rm{II}}}(\tilde{y}) \, dA_{\rm{ell}} (\tilde{y}) }
\label{edII}
\end{equation}

\noindent
with $M_{\rm{sh}}(\tilde{y})= \int_{z_d}^{z_u} \rho_{\rm{II}}(z) \, dV_{II} $ the mass inside the shell with the surface $A_{\rm{ell}}$ of the complete ellipsoid.\\
Furthermore, we have to estimate the mass inside the hot bubble interior, which is the mass ejected by the SN-explosions.  
The total mass of each star belonging to the association goes into ejecta, except $\sim 1.4\, M_{\odot}$ for a neutron star remnant, which is lower than the Oppenheimer-Volkoff limit, because most SN progenitors are 
lower mass stars among the massive stars. 
Thus, we have to consider the mass of $\hat{m} = m - 1.4$ per star with
$\hat{m}$ in units of solar masses and the total number of stars $N(m)=N(\hat{m})$. 
In order to obtain the ejecta mass for the IMF-model as a function of time (i.e. time variable $\tilde{y}$), we fix the upper mass limit -- which is related to the 
first SN-explosion -- and introduce a variable lower mass limit
$m_l( \tilde{y} )$. 
Again, connecting the mass of a star to its main sequence lifetime gives a time dependent ejecta mass.
In order to account for the mass included in the mass interval $(m_u-1,m_u)$ for integral mass bins at time $t=\tilde{y}=0$, we have to correct for the main sequence lifetime 
of stars with mass $m_u-1$ 
\begin{equation}
m_l( \tilde{y} ) = \left( \frac{t(\tilde{y}) + t(m_u-1)} {\kappa} \right)^{-\frac{1}{\alpha}} \, .
\end{equation}

\noindent
Integrating over the mass range 
of the association gives the ejecta mass of SNe exploded until some time $\tilde{y}$   

\begin{equation}
M_{\rm{ej}}( \tilde{y} ) = \int_{m_l(\tilde{y})}^{m_u} N(\hat{m}) \cdot \hat{m} \, d\hat{m} \, . 
\label{eMej}
\end{equation}

\noindent
So we are able to derive the mass ejected until the time $\tilde{y}_{\rm{rti}}$, where instabilities start to appear in the shell. 
Unfortunately, this formula does not hold for small associations, since all SNe may have exploded in a rather small time interval
$\Delta \tau = \tau_{\rm{ms}}(8) - \tau_{\rm{ms}}(m_u-1)$, possibly long before $\tilde{y}_{\rm{rti}}$. 
In that case, the ejecta mass is replaced by the total mass of the ejecta $M_{\rm{tot}}= \int_{m_l}^{m_u} N(\hat{m}) \cdot \hat{m} \, d\hat{m} $ after the last SN-explosion, 
otherwise the mass and thus the RTI-timescale would be overestimated.
Including the ejecta mass (Eq.~(\ref{eMej})) in the density $\rho_{\rm{in}}$ yields an instability timescale (Eq.~(\ref{etrti})) as a function of the most massive star 
$m_u$ in the IMF-model, but we prefer expressing it as a function of the total number of stars in an association. The relation between these 
two parameters is fitted with an approximation of the form $m_u(N_{\rm{OB}})=a_3 \cdot N_{\rm{OB}}^{b_3} + c_3$ (see Table \ref{N0NOB}) and 
has average errors of 1.2$\, \%$ over the range of $2-500$ OB-association members.\\
In the case of the SN-model, all the mass of exploding stars (except of 1.4 M$_{\odot}$ per star) is released in the initial explosion.
We find a useful power-law approximation for the total ejected mass of a star cluster as a function of the number of SNe (using an IMF with 
$\Gamma = -1.35$)
\begin{equation}
M_{\rm{tot}}(N_{\rm{SN}})= 7.98 \cdot N_{\rm{SN}}^{1.1}.
\end{equation}

\noindent
Errors for this approximation are about $3 \%$ for small associations (2 OB-stars) and $\sim 1 \%$ for large clusters (500 stars).
It can be used to obtain the ejecta mass for the wind-model as well, but instead of the number of supernovae, $N_{\rm{SN}}$, we have to include
the energy input rate into the bubble until the point of acceleration in units of the standard 
SN energy $N_{\rm{SN}} = L_w \cdot t(\tilde{y}_{\rm{acc,w}})/E_{\rm{SN}}$.
Calculating fragmentation this way only works for a wind, which is produced by averaging SNe and thus includes the mass of the exploded stars, but not for a true stellar
wind.

\begin{figure*}[t]
  \centering
  \includegraphics[angle=270,width=\textwidth, bb=60 50 544 1490, clip]{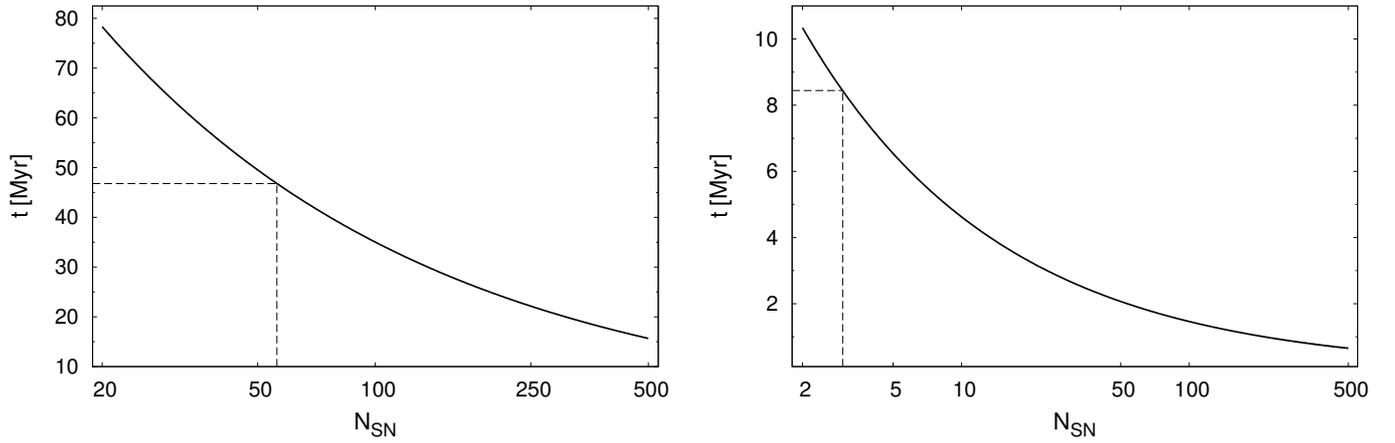}
     \caption{\label{frag_SN}Fragmentation timescales for the SN-model; the dashed line marks the
minimum number of SNe for a blow-out superbubble and the corresponding time until full fragmentation of the shell at the top of the bubble (for details see text and Tables \ref{accel} and \ref{frag}). 
Left: double exponential layer with 
$H = 500 \, $pc and $n_0=0.5\,$cm$^{-3}$ (symmetric model); 
right: exponentially stratified medium with $H = 100 \, $pc and $n_0=10\,$cm$^{-3}$ (off-plane model).}
  \end{figure*}

\begin{figure*}
  \centering
   \includegraphics[angle=270,width=\textwidth, bb=60 50 544 1490, clip]{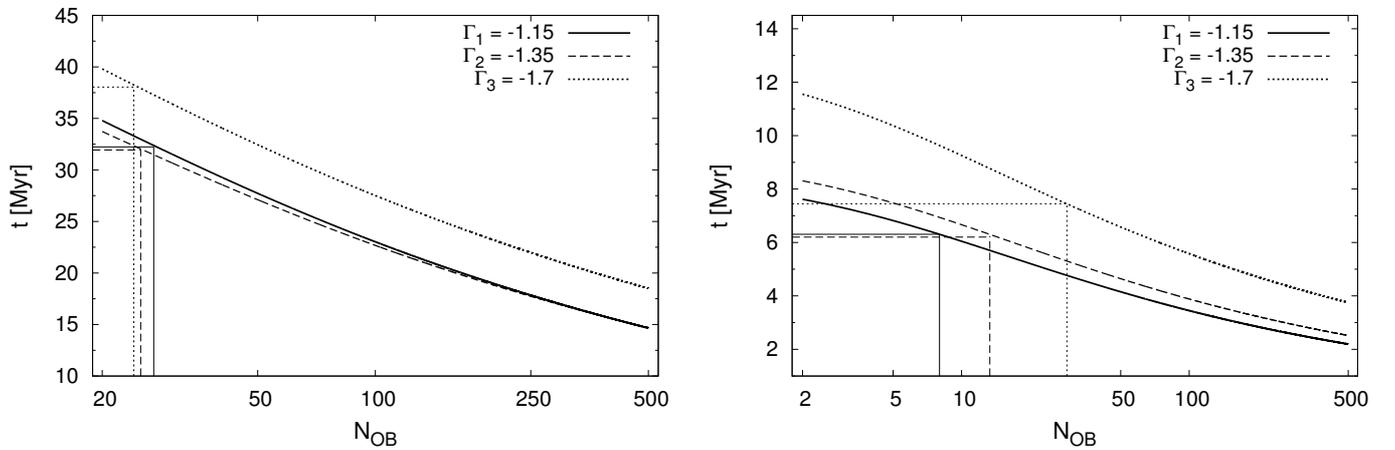}
      \caption{\label{frag_imf}Same as Fig.~\ref{frag_SN}, but for the IMF-model and indicating the minimum number of OB-stars in an association for different IMF slopes by the corresponding thin line
(left: symmetric model; right: off-plane model).}
   \end{figure*}

\begin{figure*}
  \centering
   \includegraphics[angle=270,width=\textwidth, bb=50 50 544 1490, clip]{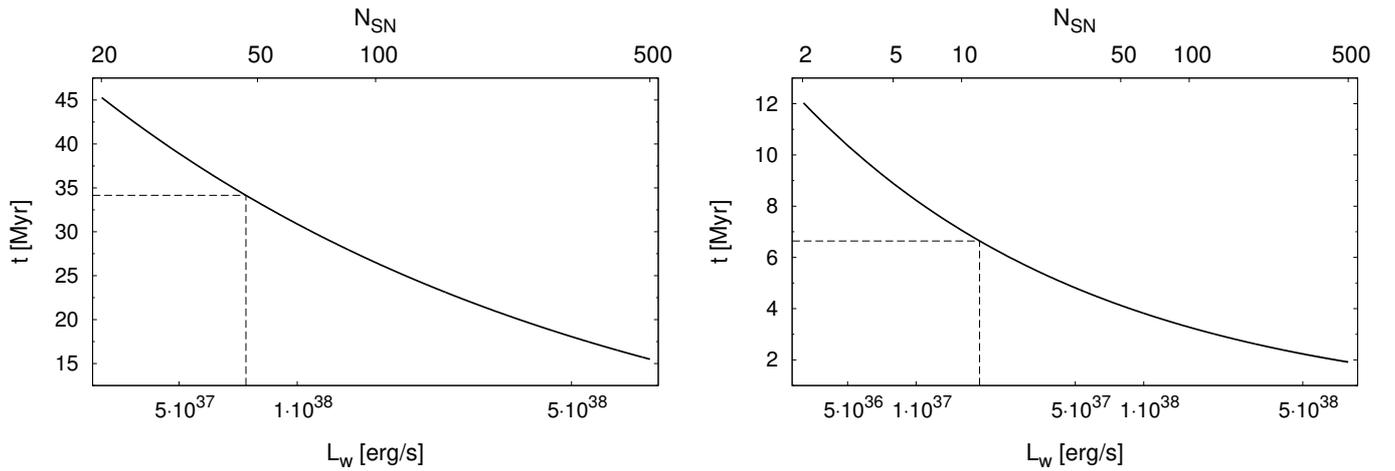}
      \caption{\label{frag_wind}Same as Fig.~\ref{frag_SN}, but for the wind model and a minimum constant energy input rate equal to a number of SNe distributed over 20 Myr
(left: symmetric model; right: off-plane model).}
   \end{figure*}

\noindent
%
Instabilities will dominate when the RTI-timescale at the top of the bubble becomes smaller than the dynamical timescale $\tau_{\rm{dyn,I}}= a/\dot{z}_u$
(symmetric bubble) or $\tau_{\rm{dyn,II}}= z_u/\dot{z}_u$ (off-plane model) of the system.
In terms of the dimensionless time variable this happens at $\tilde{y}_{\rm{rti}}$ and we have to find the value, where $\tau_{\rm{rti}}(\tilde{y}_{\rm{rti}})
\sim \tau_{\rm{dyn}}(\tilde{y}_{\rm{rti}}) $ for all models, which is shortly after the acceleration sets in. 
%
The exponentially growing instability is usually fully developed and the shell will break-up at $\tilde{y}_{\rm{frag}}$ when 
$\tau_{\rm{rti}}(\tilde{y}_{\rm{frag}}) = 1/3 \cdot \tau_{\rm{rti}}(\tilde{y}_{\rm{rti}})$ (see Table {\ref{accel}}). 
This means that the exponentially growing amplitude of the perturbation has reached a size of $e^3 \simeq 20$ times of the initial one, sufficiently large to assume full fragmentation. \\
Fig.~\ref{frag_SN} shows the fragmentation timescale as a function of the total number of stars for the SN-model, Fig.~\ref{frag_imf} shows the same for the IMF-model for different slopes
and finally, the timescale as a function of wind luminosity is presented in Fig.~\ref{frag_wind}. \\
Due to the large fragmentation timescales for very small associations, the abscissa is chosen to range from $20-500$ OB-stars for the symmetric model 
to highlight the behavior of this function. The off-plane model is shown again for $2-500$ stars. For the wind-model these numbers  
correspond to $3.2\cdot10^{37} - 7.9 \cdot 10^{38} \,$erg/s (symmetric model) and $3.2\cdot10^{36} - 7.9 \cdot 10^{38} \,$erg/s (off-plane model).\\
In both cases of ISM density distribution, fragmentation is easily achieved for clusters with $N_{\rm{blow}}$ (i.e. those producing blow-out superbubbles) within 50 Myr, a reasonable timespan for fragmentation to take place
before galactic rotation or turbulences have had a major influence on the bubble structure (see Figs.~\ref{frag_SN}, \ref{frag_imf} and \ref{frag_wind}, right panel) and no additional fragmentation criterion is needed.
However, we want to check at this point whether the fragmenting SB can escape from the disk and accelerate into the halo. Thus, we compare the acceleration of the top of the bubble\footnote{\label{foot:4}Although the bubble 
is now fully fragmented at $z_u$, we can still use the calculations from the KA since a secondary shock is formed when the hot SB interior will be ejected.} at $\tilde{y}_{\rm{frag}}$ with the gravitational acceleration 
near the galactic plane
(we do not require the bubble to escape completely from the gravitational potential of the galaxy). 
%
We use $g_z (R,z)= -\partial \phi / \partial z$ with $\phi$, the disk potential in cylindrical coordinates \citep{mn75}, given by 
\begin{equation}
\phi (R,z) = - \frac{GM}{\sqrt{R^2 + (a + \sqrt{z^2 + b^2})^2}},
\end{equation}
\clearpage
\clearpage\noindent%
where $R$ is the galactocentric radius, $G$ is the gravitational constant, $a = 7.258 \, $kpc, $b= 0.520 \,$kpc, and the disk mass $M = 2.547 \cdot 10^{11}\, $M$_{\odot}$ \citep{bmv91} for a Milky-Way-type galaxy.
This yields, e.g.~for the galactocentric distance of the Sun at $R_{\odot} =8.5 \,$kpc a value of $-g_z(R=R_{\odot}, z=1 \, \rm{pc})=3.47 \cdot 10^{-11} \,$cm s$^{-2}$.
We find for all models that SBs driven by the minimum blow-out energy, $N_{\rm{blow}} \cdot E_{\rm{SN}} $, have an acceleration larger than $-g_z(R,z=1 \, \rm{pc})$ for all galactrocentric radii $R$. 
The complete set of results can be found in Table \ref{frag}.\\
\begin{table*}
\begin{centering}
\begin{tabular}{cccccc}
\hline
\hline
Model & $\tilde{y}_{\rm{acc}}$  & $z_{u,\rm{acc}}$ & $N_{\rm{blow}}$ & $t(\tilde{y}_{\rm{acc}}$)  & $t(\tilde{y}_{\rm{acc}}$)/$t_D$  \\   [0.02cm]
      &                         & [H]              &                 &  [Myr]                     &   \\ 
\hline
SN$_{\rm{sym}}$                        & 1.380  & 2.34 &        56            &   38.9      &      2.54              \\
Wind$_{\rm{sym}}$ ($\Gamma_0=-0.932$)  & 0.963  & 1.31 &   47 (7.4 10$^{37}$) &   26.1      &      1.62             \\
IMF$_{\rm{sym}}$ ($\Gamma_1=-1.15$)    & 0.858  & 1.12 &        27            &   22.8      &      1.41               \\
IMF$_{\rm{sym}}$ ($\Gamma_2=-1.35$)    & 0.760  & 0.96 &        25            &   20.1      &      1.23           \\
IMF$_{\rm{sym}}$ ($\Gamma_3=-1.7$)     & 0.583  & 0.69 &        24            &   15.2      &      0.93              \\
\hline
SN$_{\rm{off}}$                        & 1.325  & 2.87 &       3              &    7.2      &      1.93          \\
Wind$_{\rm{off}}$ ($\Gamma_0=-0.932$)  & 0.886  & 1.87 &  12 (1.9 10$^{37}$)  &    4.7      &      1.30        \\
IMF$_{\rm{off}}$ ($\Gamma_1=-1.15$)    & 0.779  & 1.69 &       8              &    4.0      &      1.14          \\
IMF$_{\rm{off}}$ ($\Gamma_2=-1.35$)    & 0.680  & 1.53 &       13             &    3.5      &      0.99          \\
IMF$_{\rm{off}}$ ($\Gamma_3=-1.7$)     & 0.509  & 1.29 &       29             &    2.6      &      0.74         \\
\hline                                                                    
\end{tabular}                                                                          
\caption{\label{accel} Characteristic values of superbubble blow-out: rows $1-5$ of the table: symmetric model in a Lockman layer with $H=500 \,$pc, $n_0 = 0.5 \, \rm{cm}^{-3}$;
rows $6-10$: off-plane model with the position of the star cluster at $z_0=0.7 \,$H and $H = 100 \,$pc, $n_0 = 10 \, \rm{cm}^{-3}$. 
$\tilde{y}_{\rm{acc}}$: beginning of acceleration at $z_u$ (top of the bubble) in terms of $\tilde{y}$; 
$z_{u,\rm{acc}}$: coordinate $z_u$ at $\tilde{y}_{\rm{acc}}$; 
$N_{\rm{blow}}$: minimum number of stars for blow-out, the wind luminosity in erg/s is transformed into a number of SNe using a time interval of 20 Myr;
$t(\tilde{y}_{\rm{acc}}$): time at acceleration in units of Myrs for a bubble with $N_{\rm{blow}}$ stars;
$t(\tilde{y}_{\rm{acc}}$)/$t_D$: time at acceleration in units of the characteristic timescale ($t_D$ means $t_{\rm{SN}}$, Eq.~(\ref{tSN}) or $t_{\rm{IMF}}$, Eq.~(\ref{etIMF}), for the respective model).}
\end{centering}
\end{table*}

\begin{table*}
\begin{centering}
\begin{tabular}{ccccccccccc}
\hline
\hline
Model &  $\tilde{y}_{\rm{rti}}$ & $z_{u,\rm{rti}}$ & $t(\tilde{y}_{\rm{rti}}$) & $t(\tilde{y}_{\rm{rti}}$)/$t_D$ & $\ddot{z}_{u,\rm{rti}}$ & $\tilde{y}_{\rm{frag}}$ & $z_{u,\rm{frag}}$  & $t(\tilde{y}_{\rm{frag}}$) & 
$t(\tilde{y}_{\rm{frag}}$)/$t_D$ & $\ddot{z}_{u,\rm{frag}}$ \\   [0.02cm]
      &                         &     [H]          &        [Myr]                  &                                 &     [cm s$^{-2}$]       &                         &      [H]           &        [Myr]               & 
                                 &    [cm s$^{-2}$]  \\ 
\hline
SN$_{\rm{sym}}$                          &   1.389  &  2.37   &     39.7   &   2.59  &   1.7 10$^{-11}$    &   1.469       &  2.65   &     46.8   &   3.05    &   1.7 10$^{-10}$      \\
Wind$_{\rm{sym}}$ ($\Gamma_0=-0.932$)    &   0.976  &  1.34   &     26.7   &   1.67  &   2.2 10$^{-11}$    &   1.117       &  1.64   &     34.2   &   2.13    &   2.4 10$^{-10}$       \\
IMF$_{\rm{sym}}$ ($\Gamma_1=-1.15$)      &   0.873  &  1.15   &     23.4   &   1.46  &   2.7 10$^{-11}$    &   1.056       &  1.50   &     32.2   &   2.00    &   3.1 10$^{-10}$        \\
IMF$_{\rm{sym}}$ ($\Gamma_2=-1.35$)      &   0.777  &  0.98   &     20.1   &   1.27  &   3.2 10$^{-11}$    &   1.021       &  1.43   &     31.9   &   1.95    &   4.1 10$^{-10}$       \\
IMF$_{\rm{sym}}$ ($\Gamma_3=-1.7$)       &   0.604  &  0.72   &     15.9   &   0.97  &   4.5 10$^{-11}$    &   1.126       &  1.66   &     38.0   &   2.32    &   8.7 10$^{-10}$        \\
\hline
SN$_{\rm{off}}$                          &   1.333  &  2.90   &     7.3    &   1.96  &   8.6 10$^{-11}$    &   1.404       &  3.12   &     8.4    &   2.26    &   7.9 10$^{-10}$    \\
Wind$_{\rm{off}}$ ($\Gamma_0=-0.932$)    &   0.902  &  1.90   &     4.8    &   1.34  &   1.5 10$^{-10}$    &   1.084       &  2.26   &     6.6    &   1.85    &   1.6 10$^{-9}$    \\
IMF$_{\rm{off}}$ ($\Gamma_1=-1.15$)      &   0.797  &  1.72   &     4.2    &   1.18  &   1.8 10$^{-10}$    &   1.029       &  2.15   &     6.3    &   1.78    &   2.1 10$^{-9}$     \\
IMF$_{\rm{off}}$ ($\Gamma_2=-1.35$)      &   0.700  &  1.56   &     3.6    &   1.03  &   2.2 10$^{-10}$    &   1.004       &  2.09   &     6.2    &   1.77    &   2.8 10$^{-9}$     \\
IMF$_{\rm{off}}$ ($\Gamma_3=-1.7$)       &   0.531  &  1.32   &     2.8    &   0.79  &   2.9 10$^{-10}$    &   1.109       &  2.32   &     7.4    &   2.12    &   5.4 10$^{-9}$    \\
\hline                                                                    
\end{tabular}                                                                          
\caption{\label{frag} Characteristic values of superbubble fragmentation (rows $1-10$ and numbers of $N_{\rm{blow}}$ are the same as in Table \ref{accel}). 
$\tilde{y}_{\rm{rti}}$: onset of Rayleigh-Taylor instabilities (RTIs) at $z_u$; 
$z_{u,\rm{rti}}$: coordinate $z_u$ at $\tilde{y}_{\rm{rti}}$;
$t(\tilde{y}_{\rm{rti}}$): time at onset of RTIs in units of Myrs for a bubble with $N_{\rm{blow}}$ stars;
$t(\tilde{y}_{\rm{rti}}$)/$t_D$: time at onset of RTIs in units of the characteristic timescale $t_D$; 
$\ddot{z}_{u,\rm{rti}}$: acceleration of the top of the bubble at $\tilde{y}_{\rm{rti}}$ for $N_{\rm{blow}}$ stars;
$\tilde{y}_{\rm{frag}}$: full fragmentation at $z_u$; 
$z_{u,\rm{frag}}$: coordinate $z_u$ at $\tilde{y}_{\rm{frag}}$;
$t(\tilde{y}_{\rm{frag}}$): time at fragmentation in units of Myrs for a bubble with $N_{\rm{blow}}$ stars;
$t(\tilde{y}_{\rm{frag}}$)/$t_D$: time at fragmentation in units of the characteristic timescale $t_D$; 
$\ddot{z}_{u,\rm{frag}}$: acceleration of the top of the bubble at $\tilde{y}_{\rm{frag}}$ for $N_{\rm{blow}}$ stars
.}
\end{centering}
\end{table*}

\section{Application to the W4 superbubble}
%
%
%
%
Winds of young massive stars of the cluster OCl 352 are supposed to be the energy source of the W4 superbubble.
The cluster is located at a height of $\sim 35 \,$pc \citep{dts97} above the disk at a distance of $\sim 2.35 \,$ kpc \citep{w07}.
In the following, we want to apply our off-plane wind-model to this superbubble. \\ 
According to \citet{w07} only the structure above OCl 352 should be termed superbubble or chimney (G134.4+3.85), while the lower part is the W4 loop. 
The superbubble is in the process of evolving into a chimney, because the ionized shell of G134.4+3.85 already started to fragment at the top of the bubble, 
where the shell is expected to break-up first due to instabilities.
From HI observations they derive a scale height of the ambient ISM of $140 \pm 40 \,$pc and they get bubble coordinates of $z_{u,II}(\tilde{y}) \cong 246 \,$pc (i.e.
$z_{u,I}(\tilde{y}) \cong 211 \,$pc) and $r_{\rm{max}}(\tilde{y}) \cong 82 \,$pc.  With this ratio of $z_{u,I}/r_{\rm{max}} =2.57$ the bubble has reached an
evolutionary parameter of $\tilde{y}=1.93$ in the Kompaneets model. At this time, the bubble's extension from the star cluster to the top of the bubble is $6.66 \,$H 
and from the cluster to the bottom it is about $-1.35 \,$H.
Comparing this to the physical dimensions yields a scale height of only $\sim 32\,$pc in this region of the Milky Way. The scale height 
of $140 \,$pc cannot be confirmed by the KA, since this 
would result in a value for $z_d$ of almost 190 pc, i.e. the bottom of the shell reaching 150 pc below the midplane.
Our model predicts that $z_d$ should have reached only $\sim 8 \,$pc below the Galactic plane. The cluster itself is found at $z_0=1.09\,$H.
We obtain an age of the SB of 1.8 Myr or 2.3 Myr, taking a density of $n_0=5\,$cm$^{-3}$ and
$n_0=10\,$cm$^{-3}$, respectively, and using the energy input rate of $3 \cdot 10^{37} \,$erg/s (Normandeau et al. 1996).
%

%
\begin{figure*}
\centering
\includegraphics[width=0.95\textwidth]{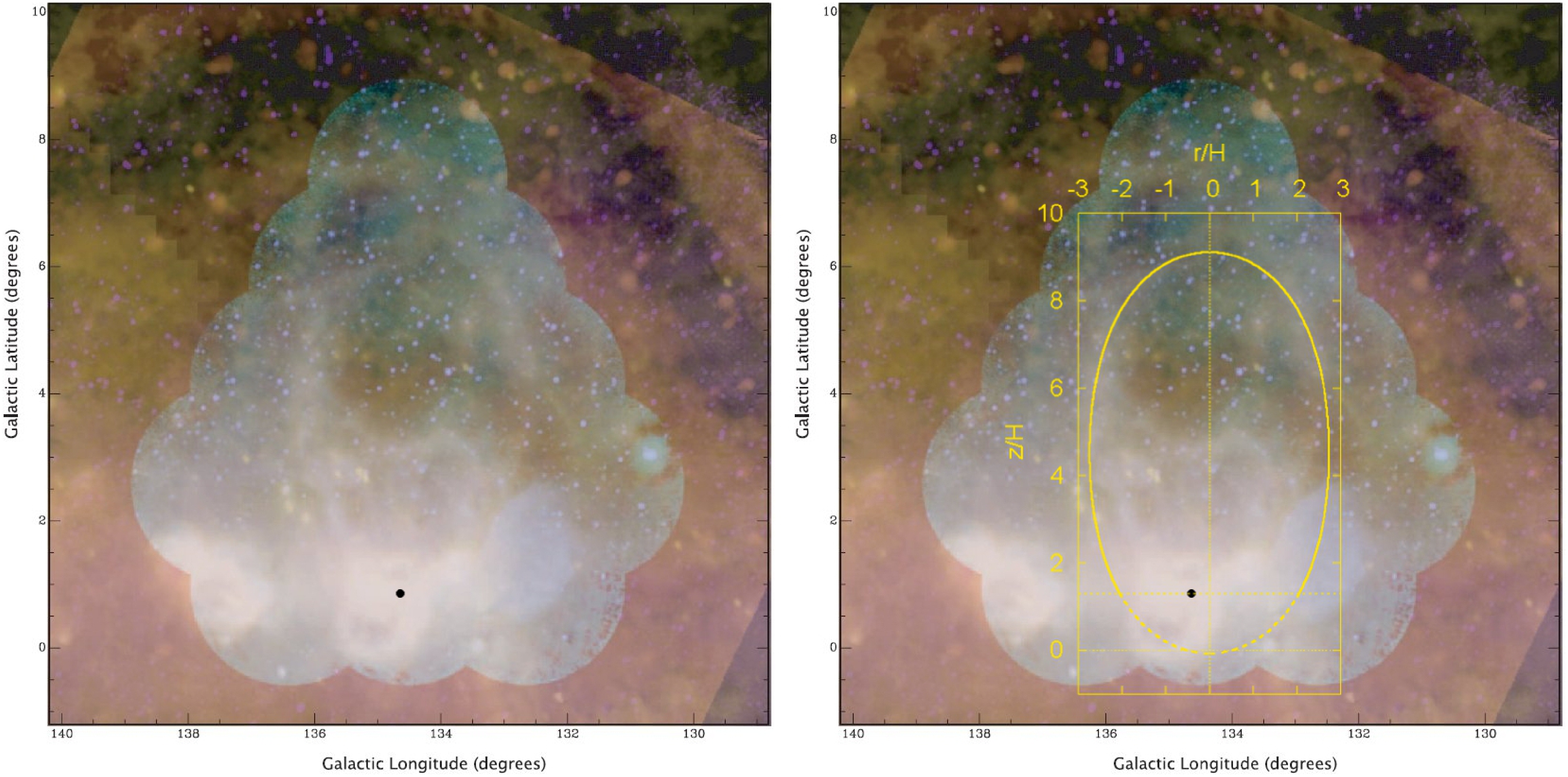}
\caption{\label{W4} Left: color image of W4 (\citet{w07}) combining HI, H$\alpha$ and infrared data. For comparison with our model, due to 
the faint structures and low contrast, the overlay is shown in a separate figure (right panel). 
The height of the star cluster above the plane is $z_0=35 \,$pc, marked by the black dot. Right: Kompaneets bubble at $\tilde{y} =1.96$ overlaid on the same image.  
The upper part of the contour fits quite well (solid line), while the part of the shell below the cluster is not seen in the observations (dotted line).}
\end{figure*}

\noindent
According to \citet{w07}, Basu et al.'s (1999) way of treating the W4 SB and W4 loop as one entity 
(and fitting a Kompaneets model to that) seems inappropriate. However, we wanted to see if our model could improve their findings.
Since \citet{bjm99} do not use an off-plane model, we first checked their results with our model for an explosion in the midplane. 
We infer an age of $\sim 2.3\,$ Myr ($\tilde{y}=1.98$) when inserting all the values they use, which corresponds quite well
to their derived age of 2.5 Myr. In this case, the distance of the top of the bubble to the star cluster of $z_1=246 \,$pc was used, which 
corresponds to the coordinate $z_{u,II}(\tilde{y})$ in the off-plane model including the offset $z_0$.
In fact, the coordinate $z_{u,I}=211\,$pc should be used as the distance of the top of the bubble to the cluster when determining the
aspect ratio, which is $z_{u,I}/r_{\rm{max}} =211/74=2.85$. With a slightly larger elongation than in the previous calculation, the bubble  
has reached the time parameter $\tilde{y} =1.96$.
With a height of $7.81 \,$H from $z_0$ to $z_{u,II}$, this yields a scale height of $H \cong 27\,$pc close to Basu et al.'s value
of 25 pc.
But, the fact that the bubble is shifted to a lower density environment changes the age of the bubble significantly. With an offset of $z_0=1.30\,$H, the age of the 
bubble is only $\sim 1.7 \,$Myr using $n_0=10\,$cm$^{-3}$. This age is found within the previous estimates for the cluster of $1.3-2.5 \,$Myr 
(cf. \citet{dts97} and references therein) 
and is therefore consistent with the assumption that the bubble is blown by the wind of the O-stars in the cluster.\\ 
%
As the W4 loop and the W4 bubble are not dynamically connected, we concentrate in our approach on fitting only the SB shell above the cluster 
with our off-plane model to Fig.~10 of \citet{w07}. 
If we use $\tilde{y}=1.93$ and the coordinates given in \citet{w07} we find that the contour of the model does not match the shell in the observations very well.
The model would fit almost perfectly, if it was shifted upwards by 
about one scale height, but then the position of the star cluster would be located outside the contour like in  
Fig.~11 of \citet{w07}.
The somewhat more elongated bubble at $\tilde{y} =1.96$ respresents nicely the shell above OCl 352 (Fig.~\ref{W4}, right) 
with the cluster -- although not exactly in the center -- matching the offset $z_0$ in the observations.
We suggest that the part of the shell below OCl 352 (Fig.~\ref{W4}, right, dashed line) was decelerated due to the presence of the W4 loop and appears now 
flattened or has even merged with the upwards expanding part of the W4 loop.\\
Applying our criterion from Sect.~3, we find that blow-out of a bubble with the association  
at around one scale height in an interstellar medium with $H=27 \,$pc and $n_0=10\,$cm$^{-3}$ is guaranteed for a wind luminosity as low as $8 \cdot 10^{35}\,$erg/s. 
We thus support Basu et al.'s statement that the bubble is 
already on its way of blowing out into the Galactic halo. 
Even if the bubble was not shifted above the plane, around $4 \cdot 10^{36}\,$erg/s would be sufficient for blow-out, which is well below the 
$1.67 \cdot 10^{38}\,$erg/s found by MMN89 (as cited in \citet{w07}).
The acceleration of the bubble, i.e. blow-out, has started already at $\tilde{y}=0.886$, more than 1 Myr 
ago.
%
%
%
%
%
%
\subsection{Local Bubble}
\begin{figure*}
\centering
\includegraphics[width=0.48\textwidth, bb=64 77 709 848, clip]{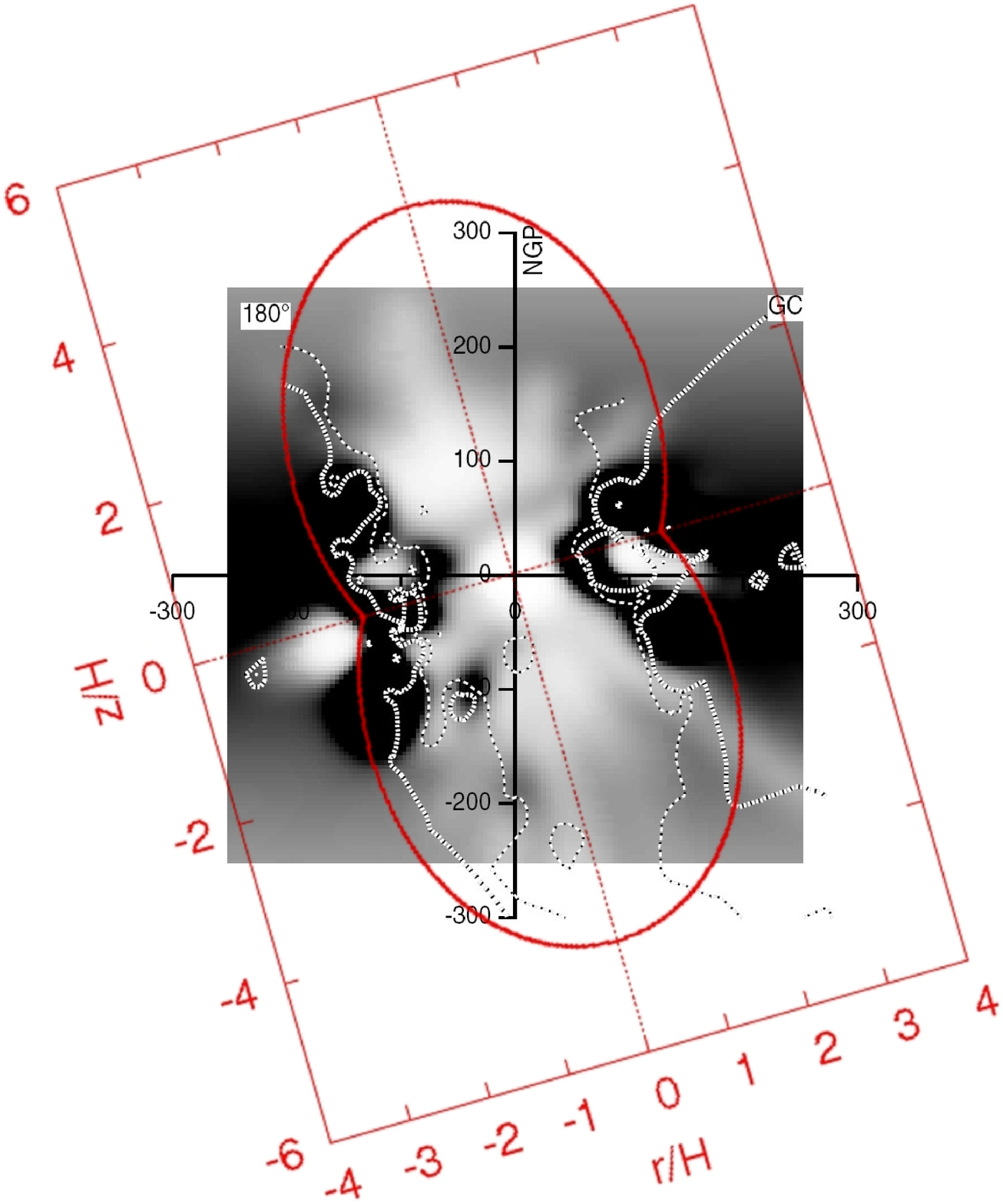}
\caption{\label{LB} Symmetric Kompaneets bubble at $\tilde{y} =1.8$ overlaid on the absorption map of neutral interstellar NaI of \citet{l03}. 
Isodensity contours with an equivalent width of 20 m$\mathring{\rm{A}}$ (inner contour) and 50 m$\mathring{\rm{A}}$ (outer contour) represent the rarefied cavity.}
\end{figure*}

\noindent
The Local Bubble (LB) was most likely produced by $14-20$ SNe, which exploded $10-15$ Myr ago as the Pleiades subgroup B1 was moving through our Galactic neighbourhood
\citep{bb02,f06}. 
Numerical simulations suggest that 19 SNe are responsible and the bubble is $\sim 13.6 \,$Myr old with the last SN having exploded 0.5 Myr ago \citep{ba06,ab12}.
Observations \citep{wssl99,slcw99} show that the bubble is not confined at higher galactic latitudes and thus, should be termed "Local Chimney", but an elongated structure (the chimney walls) 
still exists, extending to $\sim 250 - 400 \,$pc above and below the midplane \citep{l03}.
Although the Local Bubble is tilted about 20$^{\circ}$ to the midplane and expands perpendicularly to the plane of Gould's Belt, we apply here our symmetric 
model with a time-dependent energy input rate. 
%

The walls of the Local Bubble (Fig.~5 of \citet{l03}) can be fit very well with a Kompaneets model using an evolutionary parameter of $\tilde{y} =1.8$.
With the coordinate $z_u(\tilde{y}) = 4.6 \,$H corresponding to $\sim 370 \,$pc,
this yields a scale height of $H=80 \,$pc. The radius of the bubble in the plane is almost 150 pc and $r_{\rm{max}}$ has an extension of
180 pc at $\tilde{y} =1.8$ in our model, corresponding to the observations (see Fig.~\ref{LB}). 
Since it was suggested that 19 SNe exploded in a time interval of 13.1 Myr and with a given lower mass limit of $m_l=8.2 \, $M$_{\odot}$ \citep{f06}, 
we can infer -- using main sequence lifetimes of the stars -- that the upper mass boundary should be $m_u = 20.9 \, $M$_{\odot}$ (independent of the IMF). 
So far, we assumed that there is exactly one star in the mass bin ($m_u-1, m_u$) for general modeling. But this way of mass binning has to be modified and
the mass interval of the most massive star needs to be adopted, because we know both $m_u$ and $N_{\rm{OB}}$ in the case of the LB.
Using an IMF with $\Gamma_2=-1.35$, the mass bin containing the most massive star is $N(m_u-1.9,m_u)=1$ and a normalization constant of $N_0= 613$ is found. 
This simply means, that the remaining 18 stars are distributed within the mass interval from $8.2 - 19 \, $M$_{\odot}$.
A density of the undisturbed ISM of $n_0 \approx 7 \,$cm$^{-3}$ is obtained to infer the presumable age of the LB. 
Acceleration of the shell started already at $\tilde{y} = 0.760$ in this configuration, which was 3.3 Myr after the first SN exploded. The velocity of the bubble at this 
time was about 20 km/s; thus the bubble fulfills the Kompaneets criterion of blow-out into the halo. 
Rayleigh-Taylor instabilities started to appear at the top of the bubble at $\tilde{y} = 0.777$, about 3.4 Myr after the initial SN-explosion
and full fragmentation took place at a time of 5.2 Myr. With an acceleration of $\sim 2.5 \cdot 10^{-9}\,$cm s$^{-2}$ at $\tilde{y}_{\rm{frag}}$ exceeding the gravitational acceleration near the galactic plane by two orders of magnitude, 
the fragmenting shell will not fall back onto the disk. Actually, it is even higher than the local vertical component of the gravitational acceleration $-g_z(R=8.5 \,$kpc, $z=114 \, \rm{pc})=2.37 \cdot 10^{-9}$cm s$^{-2}$ such that 
blow-out of the Galaxy's gravitational potential could be achieved.
But since the LB is already disrupted on its poles before having reached one scale height of the Lockman 
layer, the bubble will not be able to expand to high-$z$ regions, but the ejected material will probably mix with the lower halo gas, leaving behind a "Local Chimney".

%
\section{Discussion}
%
%
%
\subsection{Comparisons of the models: blow-out}
%
\begin{figure*}
\centering
\includegraphics[angle=270,width=\textwidth]{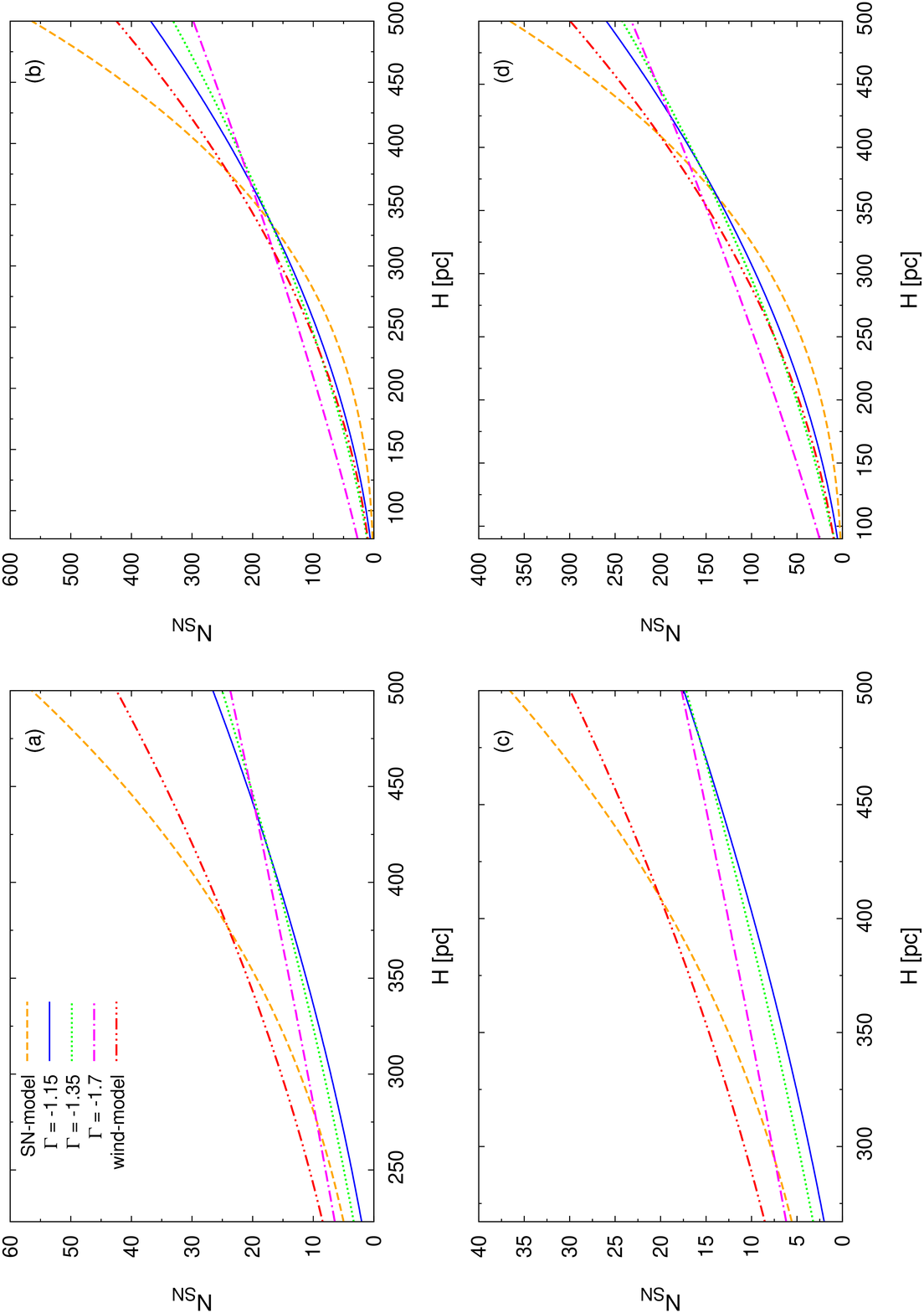}
\caption{\label{vergl} Numbers of SNe/OB-stars needed for blow-out are compared for SN-, IMF- and wind-model (energy input rate over 20 Myr gives the number of SNe
for the wind-model). 
(a) symmetric case, midplane number density: 0.5 cm$^{-3}$, scale height range: $220-500$ pc;
(b) symmetric case, midplane number density: 5 cm$^{-3}$, scale height range: $80-500$ pc;
(c) off-plane explosion at $z_0=0.7 \,$H, midplane number density: 1 cm$^{-3}$, scale height range: $270-500$ pc;
(d) off-plane explosion at $z_0=0.7 \,$H, midplane number density: 10 cm$^{-3}$, scale height range: $90-500$ pc.
}
\end{figure*}

\noindent
%
In order to compare the efficiencies of the three different models, we have to take a look at the number of SNe needed for blow-out and at the 
corresponding timescales. \\
For an ISM with properties of the Lockman layer of the Galaxy, the SN-model requires the largest number of SNe for blow-out. 
If taking into account an OB-associaton's lifetime of 
20 Myr, the wind model's efficiency is between the IMF- and the SN-model. The IMF-model is the most efficient one and 
needs only half the number of SNe to explode over
the whole lifetime of the SN-model. In fact, the IMF with a slope of $\Gamma_3=-1.7$ needs the lowest number of OB-stars to produce a blow-out SB, followed by 
the one with $\Gamma_2=-1.35$ and $\Gamma_1=-1.15$, but the numbers are approximately the same (see Table \ref{accel}). Additionally, 
the blow-out-timescales for the IMF-models are shortest with a steeper slope yielding a shorter timescale.\\
For the low scale height-high density ISM used in combination with an off-plane explosion, the 
numbers of $N_{\rm{blow}}$ are lower in general. The IMF-model with $\Gamma_3=-1.7$ is an exception, since the number of OB-stars 
$N_{\rm{blow}}$ is now higher than in the symmetric case.
The SN-model is the most efficient one with only three SNe needed for blow-out, and also
the behavior of the IMF-slopes is the opposite as above. The minimum energy input for the wind-model is somewhere in between, 
similar to that of the IMF-model with $\Gamma_2$.  
But the timescales for blow-out are unaffected by this change and are still longest for the SN-model and shortest for the $\Gamma_3$-model.
This is simply because the acceleration starts later in terms of the time variable $\tilde{y}$ for the SN-model, 
thus a larger volume has to be carved out by the SN-explosions.\\
We find the mathematical explanation for the blow-out efficiency of the models when looking at Eqs.~(\ref{eNSN}), (\ref{eN0min}), and (\ref{eLwmin}), 
which are used to describe $N_{\rm{SN,\, min}}$, $N_{\rm{OB,\, min}}$ and $L_{\rm{w,\, min}}$, respectively.
The dependence on the density is linear for the SN- and wind-model and slightly increasing with an 
exponent $b_2$ from Table \ref{N0NOB} for the IMF-models.
The SN-model is strongly influenced by the scale height, since $N_{\rm{SN,\, min}} \propto H^3$. For the wind-model, $L_{\rm{w,\, min}} \propto H^2$ is found.
The different IMF-slopes give us the following relations: $N_{\rm{OB,\, min}} \propto H^{1.9}$ for $\Gamma_1$, $H^{1.6}$ for $\Gamma_2$, and 
finally $H^{1.2}$ for $\Gamma_3$.\\
In Fig.~\ref{vergl} we illustrate these facts by comparing the symmetric and the off-plane model for a low-density and 
high-density ISM each. 
The values were chosen such that the Lockman-layer density of $n_0=0.5\,$cm$^{-3}$, (a), and a high-density 
layer (d) with $n_0=10\,$cm$^{-3}$, were used as reference (see Sect.~3). According to that, we wanted to investigate what happens to $N_{\rm{min}}$ with 
a density 10 times higher, i.e. $n_0=5\,$cm$^{-3}$, in the symmetric case (b) or 
a 10 times lower density, i.e. $n_0=1\,$cm$^{-3}$, in the off-plane scenario (c).
The plots start with a scale height on the x-axis, where at least $N_{\rm{min}} \ge 2$ are needed, 
which is the smallest association we use
in this paper, and end at a scale height of 500 pc. We find that the models behave similarly throughout this scale height
range irrespective of the density distribution, but the order of the efficiency of all models changes at $H \sim 300-450 \,$pc, 
which marks the transition of a low-scale height to a high-scale height medium.
In the low-scale height regime (until $\sim 350 \,$pc) the wind-model is least efficient, followed by the $\Gamma_3$-model and the SN-model.
The $\Gamma_2$-IMF is slightly better and an IMF with $\Gamma_1$ needs the lowest number of SNe.
For a scale height above 400 pc, the sequence of the models is the same as the one calculated for the Lockman layer (described above),
except for the off-plane model with $n_0=1\,$cm$^{-3}$, where the IMF-models do not intersect before $\sim 500 \,$pc according to Fig.~\ref{vergl}(c).
\subsection{Comparison of the models: fragmentation}
Next, we want to compare the timescales until fragmentation and explore the efficiencies of the different models.
The energy input needed for a bubble to achieve fragmentation is given by the minimum blow-out energy, i.e. the energy required for reaching a certain velocity $v_{\rm{acc}}$ at $\tilde{y}_{\rm{acc}}$ (see Table 2).
%
The trend until the beginning of the fragmentation process (i.e. at $\tilde{y}_{\rm{rti}}$) is the same as that for blow-out for both ISM cases, in terms of $\tilde{y}$ and in absolute timescale:
the SN-model is the least efficient one and an IMF with a steep slope yields the best results (shown in Table 3).  
Also, it takes the SN-model the longest total time $t(\tilde{y}_{\rm{frag}}$) until full fragmentation occurs for bubbles driven by $N_{\rm{blow}}$. 
But the IMF-model with $\Gamma_2=-1.35$ now shows the lowest total timescale followed by $\Gamma_1=-1.15$ with no obvious correlation among the models.
We find that the SN-model is the fastest model in completing the fragmentation process -- which means from $\tilde{y}_{\rm{rti}}$ until $\tilde{y}_{\rm{frag}}$ -- and a steep IMF ($\Gamma_3$) yields the highest values of this timespan.
Since this is the opposite to the efficiency sequence until $\tilde{y}_{\rm{rti}}$, no general trend for the fragmentation timescales can be seen.\\  
%
%
%
%
%
Concerning the interpretation of these facts one has to look at the RT-instability timescale itself. It is decreasing more rapidly for the SN-model and a flat IMF-model, because the acceleration is increasing in a slightly shallower way than for steep IMFs. 
Thus, a value of $1/3 \, \tau_{\rm{rti}}$ is reached easier for a flat IMF than for a steep one. 
%
%
%
\section{Summary $\&$ Conclusions}
%
%
We have developed analytical models based on the Kompaneets approximation (KA) in order to derive in a fairly simple and straightforward manner the physical parameters of observed superbubbles and their ambient
medium and to gain physical insight into the blow-out phenomenon associated with star forming regions.\\
%
In this paper we have deliberately refrained from building more complex models, which e.g. include stellar wind and Wolf-Rayet wind phases, 
because we have put our focus on the important dynamical phenomena of blow-out and fragmentation of the outer shell, and their dependence on the energy input source over time. 
A more detailed description of superbubble evolution models including further stellar evolutionary phases will be the subject of a forthcoming paper.\\ 
In our work the key aspect was to work out analytically the dynamics of (unfragmented) superbubbles for different energy input modes.
We modified the KA to implement a more realistic way of energy input, i.e. modeling
the time sequence of exploding stars in an OB association by including the main sequence lifetime of the massive stars and describing the numbers per mass interval by
an initial mass function. We tested three different IMF slopes and also compared the IMF-model to a simple SN-model with an instantaneous release of energy and to a wind model
with a constant energy input rate. Two different density distributions of the ISM were applied, a symmetric medium with parameters of the Lockman layer and
a high-density, low-scale height pure exponential atmosphere with the star cluster dislocated from the galactic plane.
Velocity and acceleration of the shock front can be calculated analytically and the question how many SNe are needed for blow-out into a galactic halo can be answered.
The exact position of the outer shock in scale height units when the acceleration starts can be given.
Furthermore, the timescale for the development of Rayleigh-Taylor instabilities in a SB shell is calculated, and thus, a fragmentation timescale can be derived. 
The overall pattern shows that at larger scale heights ($H > 400 \,$pc), independent of the ISM density, the SN-model needs the highest energy input, followed by the wind-model, whereas an IMF with a steep slope
is the most efficient one. The same ranking applies to the blow-out timescales of these models (Figs.~\ref{time_SN}-\ref{time_wind} $\&$ Table \ref{accel}). 
At low scale heights ($H \sim 100 \, $pc) and moderate or high densities ($n_0 \ge 5 \,$cm$^{-3}$), the picture changes completely, i.e. the SN-model 
requires the lowest energy input and the IMF-model with $\Gamma_3=-1.7$ is the least efficient one. The explanation is that a single release of energy is more powerful in sweeping up a 
thin layer of ISM, whereas it is easier for the IMF model with 
an \emph{increasing} energy input with time ($L_{\rm{SB}} \propto t^{\delta}$, $0 < \delta < 1$, Eq.~\ref{eLSB2}) to sustain the supply of energy over a larger distance. 
When comparing fragmentation timescales (Figs.~\ref{frag_SN}-\ref{frag_wind} $\&$ Table \ref{frag}), the IMF-models exhibit the lowest values, favouring flatter IMFs with $\Gamma_1$ and $\Gamma_2$.
In terms of $\tilde{y}$ and absolute time, fragmentation happens first for the $\Gamma_2$-model for SBs driven by the minimum number of SNe for blow-out.\\ 
Still, the KA is a rather simple model. It does not account for magnetic fields, ambient pressure, inertia and evaporation of the shell.
Also a galactic gravitational field and cooling of the shocked gas inside the cavity are neglected. \\
In our model, we include galactic gravity in a rudimentary way: if a bubble fulfills the blow-out criterion and additionally the shell's acceleration at the top of the bubble at the time of fragmentation exceeds the vertical component of 
the gravitational acceleration in the disk, the SN-ejecta and fragmenting shell are expelled into the halo. We find that this is true for all bubbles created by an energy input of $N_{\rm{blow}} \cdot E_{\rm{SN}} $.\\
Further analysis of a fragmenting superbubble
would also involve to calculate the motion of the shell fragments, and
ultimately
 the dynamics of a galactic fountain like e.g. in \citet{s08}. A simple ballistic treatment of the motion of blob
fragments in a gravitational potential would, however, be too simplistic, as they would for some time experience a drag force due
to the outflowing hot bubble interior. The drag would be proportional to the ram pressure of the 
hot gas and the cross section of a blob, which would, to lowest order, be proportional to the thickness of
the fragmented shell; and finally it would depend on the geometry of the blob, which might be taken as spherical. 
However, due to compressibility, bow-shocks and head-tail structures might subsequently be formed, so that
for a realistic treatment numerical simulations would be the best choice.\\
According to MM88 cooling of the bubble interior can be neglected
for Milky Way type parameters of the ISM, but should be taken into account for smaller OB-associations or a dense and cool ISM. 
In general, cooling is not important as long as the timescale for radiative cooling is large compared to the characteristic 
dynamical timescale of the superbubble.
%
%
%
Including a magnetic field would be quite important, but this goes beyond the scope of our analytical model and is left to numerical simulations.
\citet{fmz91} find that the presence of a magnetic field could slow down the expansion of a superbubble.
\citet{s09} argue that the scale height and age of a bubble are underestimated by $\sim 50 \, \%$ when using a Kompaneets model without magnetic fields. 
However, they cannot produce such narrow superbubbles like W4 with their MHD simulations.
Moreover, 3D high resolution numerical simulations \citep{ab05} show that magnetic tension forces are much less efficient in 3D than in 2D in holding back the expanding bubble.\\
A slightly slower growth of a bubble would be also achieved by taking into account the inertia of the cold massive shell in the calculations.
MM88 find a difference of $\sim 10 \, \%$ in radius after comparison 
with the models of \citet{s85}, which neglect inertia.
Also due to the ambient pressure of the ISM superbubbles should expand more slowly as it was suggested by \citet{og04}.
In our analytical calculations we find that these effects have to be compensated when we make comparisons to observed bubbles 
by including a rather high ISM density to prevent SBs from expanding too fast. 
Furthermore, a clumpy ambient medium cannot be considered by the KA.\\
We applied our models to the W4 superbubble and the Local Bubble, both in the Milky Way. 
It is certainly not easy to compare a simple model with an observed SB, which is not isolated. In the region of the W3/W4/W5 bubbles, several complexes and clouds are found 
and multiple epochs of star formation make it difficult to distinguish, which cluster has formed which bubble at what time. However, it is most likely that 
the cluster OCl 352 is responsible for driving the evolution of the bubble \citep{w07}. 
\citet{o05} suggest that winds or SNe of previous stellar generations are responsible for earlier clearing of this region, which could
explain the low scale height of around 30 pc. Our calculations suggest that the bubble is younger than found by other authors, which 
is due to the offset of the association more than one scale height above the Galactic plane. 
Shifting to lower densities makes it easier to produce a blow-out superbubble in a shorter timescale. This is an important result and should be included in the models.\\
The Local Bubble is one of the rare cases for a double-sided bubble, 
which can be tested with our symmetric superbubble model.
From geometrical properties, we estimate an ISM scale height of $\sim 80 \,$pc. In order to reproduce size and age of the bubble correctly, 
we deduce from our models that it was an intermediate density region ($n_0 \sim 7 \,$cm$^{-3}$) before the first SN-explosion around 14 Myrs ago.
Also this place in the Milky Way is very complex (neighbouring Loop I superbubble), which can't be included in our modeling of superbubbles. \\
We conclude that blow-out energies derived in this paper are lower thresholds and might be higher if e.g. magnetic fields play a role.
Accordingly, fitting the models to observed bubbles gives an upper limit for densities of the ambient ISM prior to the first SN-explosion.
Observers are encouraged to use the model presented here for deriving important physical parameters of e.g. the energy input sources (number of OB stars, richness of cluster etc.), scale heights, dynamical time scales among other quantities. The solutions of the equations derived in detail here are easy to obtain by simple mathematical programs. 
Theorists may find it useful to compare our analytic results to high resolution numerical simulations in order to separate more complex effects, such as turbulence, mass loading, magnetic fields etc. from basic physical effects, incorporated in our model. 
%
%
%
%
%
%
\begin{acknowledgements}
VB acknowledges support from the Austrian Academy of Sciences, the University of Vienna, and the Austrian FWF, and the Zentrum f\"ur Astronomie und Astrophysik (TU Berlin) for financial help during several short term visits. 
\end{acknowledgements}
\appendix
\section{Thermal energy in the bubble}
\label{app1}
The fraction $E_{\rm{th}}$ of the total energy, which is converted into thermal energy at the inner shock, is derived.
The equation for the thermal energy inside the hot bubble is  

\begin{equation}
E_{\rm{th}}(t)= \frac{1}{\gamma - 1} \cdot P(t) \cdot V(t) \, , 
\label{a1}
\end{equation}

\noindent
where $V(t) = \frac{4 \pi}{3} \, R^3(t)$ is the volume of a spherical remnant and $R(t)$ is the radius of the outer shock.
Since \citet{bjm99} have shown that the thermal energy in the hot bubble is 
comparable to that of a homogeneous one until late evolutionary stages, the simplification of a spherical volume is used in the calculations.
The pressure $P(t)$ is obtained by taking into account momentum conservation of the bubble shell (see, for example, Castor et al.~1975, Weaver et al.~1977)

\begin{equation}
\frac{d}{dt}\left(\frac{4 \, \pi}{3}R^3 \, \rho_0 \frac{dR}{dt}\right) = 4 \, \pi R^2 \cdot P \, 
\end{equation}

\noindent
with homogeneous ambient density $\rho_0$.
This equation is rearranged which gives

\begin{equation}
P(t)=\rho_0 \cdot \dot{R}^2 + \frac{1}{3}\rho_0 \, R \, \ddot{R} \, .
\label{a3}
\end{equation}

\noindent
Combining this equation with Eq.~(\ref{a1}) for the thermal energy results in

\begin{equation}
E_{\rm{th}}=3/2 \, \rho_0 \cdot \Bigl(\dot{R}^2 + \frac{1}{3} \, R \, \ddot{R}\Bigr) \cdot V \, .
\label{a4}
\end{equation}

\subsection{SN-model}
In the case of a bubble created by a single explosion, the radius of the outer shock is \citep[e.g.][]{cc07}

\begin{equation}
R_{\rm{SN}}(t) = \left( \frac{50}{9 \, \pi} \right)^{1/5} \left( \frac{E_{\rm{SN}}\cdot N_{\rm{SN}}}{\rho_0} \right)^{1/5}\, t^{2/5}.
\label{a5}
\end{equation}

\noindent
Inserting $R_{\rm{SN}}(t)$ and its derivatives $\dot{R}$ and $\ddot{R}$ to equation (\ref{a4}) results in 
$
E_{\rm{th,\, SN}} = 2/3 \cdot E_{\rm{SN}}\cdot N_{\rm{SN}} .
$

\subsection{IMF-model}
In order to describe the evolution of a superbubble driven by a time-dependent energy input rate -- in addition to the equations above -- 
energy conservation of the hot wind gas has to be considered for this problem 

\begin{equation}
\frac{dE_{\rm{th}}}{dt}= L_{\rm{SB}} - P \cdot \frac{d}{dt} \left( \frac{4 \, \pi }{3}R^3 \right)
\label{a6}
\end{equation}

\noindent
with the total thermal energy in this region given by Eq.~(\ref{a1}).
$L_{\rm{SB}}= L_{\rm{IMF}} \cdot t^{\, \delta}$ is the energy input rate delivered by sequential SN-explosions
according to an Initial Mass Function.
Additionally it is assumed that the radius of the bubble in a self-similar flow scales with time like $R_{\rm{IMF}}(t) = A_s \cdot t^{\, \mu}$.
%
%
%
The constant is 

\begin{equation}
A_s = \left( \frac{375}{(7\delta +11)(4\delta + 7)(\delta +3)} \right)^{1/5}\cdot \left( \frac{L_{\rm{IMF}}}{2\, \pi \rho_0}\right)^{1/5}
\end{equation}

\noindent
and the exponent $\mu = \frac{\delta +3}{5}$ \citep[e.g.][]{bb02}.
Now, the fraction $E_{\rm{th}}(t)$ of the total energy, which is converted into thermal energy at the inner shock, can be calculated.
Inserting $R_{\rm{IMF}}(t)$ and its derivatives into Eq.~(\ref{a4}) yields
$
E_{\rm{th,\, IMF}}(t) =\frac{5}{7 \delta + 11} \cdot L_{\rm{IMF}} \cdot t^{\, \delta + 1}.
$
\section{Acceleration of the top of the bubble}
\label{app2}
The derivative of the bubble volume with respect to $\tilde{y}$ is needed in the following calculations, which is the same for all 
models presented below.\\
We use the derivative of the series expansion of the volume $V_{\rm{I}}$ in the symmetric case.
For the offcenter model we find

\begin{equation}
\begin{split}
\frac{d\tilde{V}_{\rm{II}}(\tilde{y})}{d\tilde{y}} &=  \frac{16 \pi H^3}{3} \cdot \frac{\arcsin\frac{\tilde{y}}{2}}{(1 - \tilde{y}/2)^2} \cdot \\
&\cdot \left( \arcsin\frac{\tilde{y}}{2} + \ln \left( \frac{1 + \tilde{y}/2}{1-\tilde{y}/2}\right) \cdot \sqrt{1 - \frac{\tilde{y}^2}{4}} \right) \, ,
\end{split}
\end{equation}

\subsection{SN-model}
In order to obtain the acceleration at the top of the bubble, we need to calculate the first term on the RHS of Eq.~(\ref{eddotzu}) which is the derivative of Eq.~(\ref{edzuSN}) 
with respect to $\tilde{y}$

\begin{equation}
\begin{split}
\frac{d\dot{z}_u}{d\tilde{y}} &= \frac{H\cdot \beta}{t_{\rm{SN}}} \cdot \Bigl( \frac{1}{2} \cdot \frac{1}{(1 - \tilde{y}/2)^2} \cdot \frac{1}{\tilde{V}_{\rm{I,II}}(\tilde{y})} - \\
& - \frac{1}{2} \cdot \frac{H}{1 - \tilde{y}/2}  \cdot \frac{1}{ \tilde{V}_{\rm{I,II}}^{3/2}(\tilde{y}) } \cdot \frac{d\tilde{V}_{\rm{I,II}}(\tilde{y})}{d\tilde{y}} \Bigr) \, .
\end{split}
\end{equation}

\noindent
Multiplying the above equation by $\frac{d\tilde{y}}{dt}$ (Eq.~(\ref{edydtSN})) yields the acceleration (in units of cm/s$^2$)

\begin{equation}
\begin{split}
\ddot{z}_u(\tilde{y}) &= 
\frac{H}{t_{\rm{SN}}^2} \cdot \frac{\beta^2}{2\cdot(1-\tilde{y}/2)} \cdot \frac{1}{\tilde{V}_{\rm{I,II}}(\tilde{y})} \cdot\\ 
& \cdot\left(\frac{1}{1 - \tilde{y}/2}
- \frac{1}{\sqrt{\tilde{V}_{\rm{I,II}}(\tilde{y})}} \cdot \frac{d\tilde{V}_{\rm{I,II}}(\tilde{y})}{d\tilde{y}} \right) \, .\\
\end{split}
\end{equation}
\subsection{IMF- and wind-model}
Next, we want to determine the acceleration at the top of the bubble $\ddot{z}_u= \frac{d\dot{z}_u}{dt}$ for the IMF-model. Using $\delta=0$ gives the acceleration for the 
wind-model.
Already, $\frac{d\tilde{y}}{dt}$ is known (Eq.~(\ref{edydtimf2})),
therefore only $\frac{d\dot{z}_u}{dy}$ has to be calculated. Determining the derivative of Eq.~(\ref{edzudtimf}) with respect
to $\tilde{y}$ yields

\begin{equation}
\begin{split}
\frac{d\dot{z}_u}{d\tilde{y}} &= \frac{H}{t_{\rm{IMF}}} \cdot \xi \cdot \Bigl[\frac{1}{2} \cdot \frac{1}{(1 - \tilde{y}/2)^2} \cdot \frac{1}{\tilde{V}^{1/2}(\tilde{y})} \cdot \left(\int_0^{\tilde{y}} \sqrt{\tilde{V}(\tilde{y}')}\, d\tilde{y}' \right)^{\bar{\delta}} - \\
&-\frac{1}{2} \cdot \frac{1}{1 - \tilde{y}/2} \cdot\frac{1}{ \left(\tilde{V}(\tilde{y}) \right)^{3/2}} \cdot \frac{d\tilde{V}(\tilde{y})}{d\tilde{y}} \cdot \left(\int_0^{\tilde{y}} \sqrt{\tilde{V}(\tilde{y}')}\, d\tilde{y}' \right)^{\bar{\delta}} + \\
&\frac{1}{1 - \tilde{y}/2} \cdot \frac{1}{\tilde{V}^{1/2}(\tilde{y})} \cdot \bar{\delta} \cdot \left(\int_0^{\tilde{y}} \sqrt{\tilde{V}(\tilde{y}')}\, d\tilde{y}'\right)^{\bar{\delta} - 1}  \tilde{V}^{1/2}(\tilde{y}) \Bigr] \, .
\end{split}
\end{equation}

\noindent
For the complete expression of the acceleration $\ddot{z}_u(\tilde{y})$ in the IMF-model, the equation above just has to be multiplied by $d\tilde{y}/dt$ 
(Eq.~(\ref{edydtimf2})).
\section{List of variables \& parameters}
\label{app3}
\begin{tabbing}
$\gamma$ \quad \quad \, \= ... \quad \=  ratio of specific heats ($\gamma=5/3$)\\
$\Gamma$ \> ... \> slope of the IMF ($\Gamma_0=-0.932$, $\Gamma_1=-1.15$,\\
\> \>  $\Gamma_2=-1.35$, $\Gamma_3=-1.7$)\\
$H$ \> ... \> scale height\\
$L_{\rm{IMF}}$ \> ... \> IMF energy input rate coefficient\\
$n_0$ \> ... \> number density at galactic midplane\\
$N_0$ \> ... \> IMF normalization constant\\
$N_{\rm{blow}}$ \> ... \> minimum number of stars for blow-out\\
$\tau_{\rm{dyn}}$ \> ... \> dynamical timescale\\
$\tau_{\rm{rti}}$ \> ... \> Rayleigh-Taylor instability timescale\\
$V_I(\tilde{y})$ \> ... \> bubble volume (symmetric model)\\
$V_{II}(\tilde{y})$ \> ... \> bubble volume (off-plane model)\\
$v_{\rm{acc}}$ \> ... \> velocity at time of blow-out\\
$y$ \> ... \> transformed time variable\\
$\tilde{y}$ \> ... \> transformed time variable in scale height units\\
$\tilde{y}_{\rm{acc}}$ \> ... \> time of blow-out\\
$\tilde{y}_{\rm{frag}}$ \> ... \> time of fragmentation of the shell\\
$\tilde{y}_{\rm{rti}}$ \> ... \> time of onset of Rayleigh-Taylor instabilities\\
$z_0$ \> ... \> center of the explosion\\
$z_d$ \> ... \> bottom of the bubble\\
$z_u$ \> ... \> top of the bubble\\
$\dot{z_u}(\tilde{y})$ \> ... \> velocity of the bubble at $z_u$\\
$\ddot{z_u}(\tilde{y})$ \> ... \> acceleration of the bubble at $z_u$\\
\end{tabbing}
\bibliographystyle{aa}
\bibliography{superbubbles}

\begin{thebibliography}{64}
\expandafter\ifx\csname natexlab\endcsname\relax\def\natexlab#1{#1}\fi

\bibitem[{{Baldry} \& {Glazebrook}(2003)}]{bg03}
{Baldry}, I.~K. \& {Glazebrook}, K. 2003, \apj, 593, 258

\bibitem[{{Basu} {et~al.}(1999){Basu}, {Johnstone}, \& {Martin}}]{bjm99}
{Basu}, S., {Johnstone}, D., \& {Martin}, P.~G. 1999, \apj, 516, 843

\bibitem[{{Bergh{\"o}fer} \& {Breitschwerdt}(2002)}]{bb02}
{Bergh{\"o}fer}, T.~W. \& {Breitschwerdt}, D. 2002, \aap, 390, 299

\bibitem[{{Bisnovatyi-Kogan} \& {Silich}(1995)}]{bs95}
{Bisnovatyi-Kogan}, G.~S. \& {Silich}, S.~A. 1995, Reviews of Modern Physics,
  67, 661

\bibitem[{{Bland-Hawthorn} {et~al.}(2007){Bland-Hawthorn}, {Veilleux}, \&
  {Cecil}}]{bvc07}
{Bland-Hawthorn}, J., {Veilleux}, S., \& {Cecil}, G. 2007, \apss, 311, 87

\bibitem[{{Bregman} \& {Lloyd-Davies}(2007)}]{bl07}
{Bregman}, J.~N. \& {Lloyd-Davies}, E.~J. 2007, \apj, 669, 990

\bibitem[{{Breitschwerdt} \& {de Avillez}(2006)}]{ba06}
{Breitschwerdt}, D. \& {de Avillez}, M.~A. 2006, \aap, 452, L1

\bibitem[{{Breitschwerdt} {et~al.}(2000){Breitschwerdt}, {Freyberg}, \&
  {Egger}}]{bfe00}
{Breitschwerdt}, D., {Freyberg}, M.~J., \& {Egger}, R. 2000, \aap, 361, 303

\bibitem[{{Breitschwerdt} {et~al.}(1991){Breitschwerdt}, {McKenzie}, \&
  {Voelk}}]{bmv91}
{Breitschwerdt}, D., {McKenzie}, J.~F., \& {Voelk}, H.~J. 1991, \aap, 245, 79

\bibitem[{{Brown} {et~al.}(1994){Brown}, {de Geus}, \& {de Zeeuw}}]{b94}
{Brown}, A.~G.~A., {de Geus}, E.~J., \& {de Zeeuw}, P.~T. 1994, VizieR Online
  Data Catalog, 328, 90101

\bibitem[{{Cash} {et~al.}(1980){Cash}, {Charles}, {Bowyer}, {Walter},
  {Garmire}, \& {Riegler}}]{ccb80}
{Cash}, W., {Charles}, P., {Bowyer}, S., {et~al.} 1980, \apjl, 238, L71

\bibitem[{{Castor} {et~al.}(1975){Castor}, {McCray}, \& {Weaver}}]{cmw75}
{Castor}, J., {McCray}, R., \& {Weaver}, R. 1975, \apjl, 200, L107

\bibitem[{{Chevalier} \& {Gardner}(1974)}]{cg74}
{Chevalier}, R.~A. \& {Gardner}, J. 1974, \apj, 192, 457

\bibitem[{{Chu} \& {Mac Low}(1990)}]{cm90}
{Chu}, Y. \& {Mac Low}, M. 1990, \apj, 365, 510

\bibitem[{{Clarke} \& {Carswell}(2007)}]{cc07}
{Clarke}, C.~J. \& {Carswell}, R.~F. 2007, Principles of Astrophysical Fluid
  Dynamics, by C.~J.~Clarke and R.~F.~Carswell, pp.~226.~Cambridge University
  Press, New York

\bibitem[{{Crawford} {et~al.}(2002){Crawford}, {Lallement}, {Price}, {Sfeir},
  {Wakker}, \& {Welsh}}]{clp02}
{Crawford}, I.~A., {Lallement}, R., {Price}, R.~J., {et~al.} 2002, \mnras, 337,
  720

\bibitem[{{Dahlem} {et~al.}(1998){Dahlem}, {Weaver}, \& {Heckman}}]{dwh98}
{Dahlem}, M., {Weaver}, K.~A., \& {Heckman}, T.~M. 1998, \apjs, 118, 401

\bibitem[{{Dawson} {et~al.}(2002){Dawson}, {Spinrad}, {Stern}, {Dey}, {van
  Breugel}, {de Vries}, \& {Reuland}}]{dss02}
{Dawson}, S., {Spinrad}, H., {Stern}, D., {et~al.} 2002, \apj, 570, 92

\bibitem[{{de Avillez}(2000)}]{a00}
{de Avillez}, M.~A. 2000, \mnras, 315, 479

\bibitem[{{de Avillez} \& {Breitschwerdt}(2005)}]{ab05}
{de Avillez}, M.~A. \& {Breitschwerdt}, D. 2005, \aap, 436, 585

\bibitem[{{de Avillez} \& {Breitschwerdt}(2012)}]{ab12}
{de Avillez}, M.~A. \& {Breitschwerdt}, D. 2012, \aap, 539, L1

\bibitem[{{Dennison} {et~al.}(1997){Dennison}, {Topasna}, \&
  {Simonetti}}]{dts97}
{Dennison}, B., {Topasna}, G.~A., \& {Simonetti}, J.~H. 1997, \apjl, 474, L31

\bibitem[{{Ferrara} \& {Tolstoy}(2000)}]{ft00}
{Ferrara}, A. \& {Tolstoy}, E. 2000, \mnras, 313, 291

\bibitem[{{Ferriere} {et~al.}(1991){Ferriere}, {Mac Low}, \& {Zweibel}}]{fmz91}
{Ferriere}, K.~M., {Mac Low}, M., \& {Zweibel}, E.~G. 1991, \apj, 375, 239

\bibitem[{{Fuchs} {et~al.}(2006){Fuchs}, {Breitschwerdt}, {de Avillez},
  {Dettbarn}, \& {Flynn}}]{f06}
{Fuchs}, B., {Breitschwerdt}, D., {de Avillez}, M.~A., {Dettbarn}, C., \&
  {Flynn}, C. 2006, \mnras, 373, 993

\bibitem[{{Heiles}(1990)}]{h90}
{Heiles}, C. 1990, \apj, 354, 483

\bibitem[{{Kamphuis} {et~al.}(1991){Kamphuis}, {Sancisi}, \& {van der
  Hulst}}]{ksh91}
{Kamphuis}, J., {Sancisi}, R., \& {van der Hulst}, T. 1991, \aap, 244, L29

\bibitem[{{Kompaneets}(1960)}]{k60}
{Kompaneets}, A.~S. 1960, Soviet Phys. Doklady, 5, 46

\bibitem[{{Kontorovich} \& {Pimenov}(1998)}]{kp98}
{Kontorovich}, V.~M. \& {Pimenov}, S.~F. 1998, ArXiv Astrophysics
  e-prints:astro-ph/9802149

\bibitem[{{Korycansky}(1992)}]{k92}
{Korycansky}, D.~G. 1992, \apj, 398, 184

\bibitem[{{Lallement} {et~al.}(2003){Lallement}, {Welsh}, {Vergely}, {Crifo},
  \& {Sfeir}}]{l03}
{Lallement}, R., {Welsh}, B.~Y., {Vergely}, J.~L., {Crifo}, F., \& {Sfeir}, D.
  2003, \aap, 411, 447

\bibitem[{{Lee} \& {Chen}(2009)}]{lc09}
{Lee}, H. \& {Chen}, W.~P. 2009, \apj, 694, 1423

\bibitem[{{Lockman}(1984)}]{l84}
{Lockman}, F.~J. 1984, \apj, 283, 90

\bibitem[{{Mac Low} \& {McCray}(1988)}]{mm88}
{Mac Low}, M. \& {McCray}, R. 1988, \apj, 324, 776

\bibitem[{{Mac Low} {et~al.}(1989){Mac Low}, {McCray}, \& {Norman}}]{mmn89}
{Mac Low}, M., {McCray}, R., \& {Norman}, M.~L. 1989, \apj, 337, 141

\bibitem[{{Maciejewski} \& {Cox}(1999)}]{mc99}
{Maciejewski}, W. \& {Cox}, D.~P. 1999, \apj, 511, 792

\bibitem[{{Maciejewski} {et~al.}(1996){Maciejewski}, {Murphy}, {Lockman}, \&
  {Savage}}]{m96}
{Maciejewski}, W., {Murphy}, E.~M., {Lockman}, F.~J., \& {Savage}, B.~D. 1996,
  \apj, 469, 238

\bibitem[{{Massey}(1999)}]{m99}
{Massey}, P. 1999, in IAU Symposium, Vol. 190, New Views of the Magellanic
  Clouds, ed. {Y.-H.~Chu, N.~Suntzeff, J.~Hesser, \& D.~Bohlender}, 173

\bibitem[{{McCray} \& {Kafatos}(1987)}]{mk87}
{McCray}, R. \& {Kafatos}, M. 1987, \apj, 317, 190

\bibitem[{{Miyamoto} \& {Nagai}(1975)}]{mn75}
{Miyamoto}, M. \& {Nagai}, R. 1975, \pasj, 27, 533

\bibitem[{{Oey} \& {Garc{\'{\i}}a-Segura}(2004)}]{og04}
{Oey}, M.~S. \& {Garc{\'{\i}}a-Segura}, G. 2004, \apj, 613, 302

\bibitem[{{Oey} {et~al.}(2005){Oey}, {Watson}, {Kern}, \& {Walth}}]{o05}
{Oey}, M.~S., {Watson}, A.~M., {Kern}, K., \& {Walth}, G.~L. 2005, \aj, 129,
  393

\bibitem[{{Pidopryhora} {et~al.}(2007){Pidopryhora}, {Lockman}, \&
  {Shields}}]{pls07}
{Pidopryhora}, Y., {Lockman}, F.~J., \& {Shields}, J.~C. 2007, \apj, 656, 928

\bibitem[{{Reynolds}(1989)}]{r89}
{Reynolds}, R.~J. 1989, \apjl, 339, L29

\bibitem[{{Sakamoto} {et~al.}(2006){Sakamoto}, {Ho}, {Iono}, {Keto}, {Mao},
  {Matsushita}, {Peck}, {Wiedner}, {Wilner}, \& {Zhao}}]{s06}
{Sakamoto}, K., {Ho}, P.~T.~P., {Iono}, D., {et~al.} 2006, \apj, 636, 685

\bibitem[{{Salpeter}(1955)}]{s55}
{Salpeter}, E.~E. 1955, \apj, 121, 161

\bibitem[{{Savage} {et~al.}(1997){Savage}, {Sembach}, \& {Lu}}]{ssl97}
{Savage}, B.~D., {Sembach}, K.~R., \& {Lu}, L. 1997, \aj, 113, 2158

\bibitem[{{Scalo}(1986)}]{sca86}
{Scalo}, J.~M. 1986, \fcp, 11, 1

\bibitem[{{Schiano}(1985)}]{s85}
{Schiano}, A.~V.~R. 1985, \apj, 299, 24

\bibitem[{{Sedov}(1946)}]{s46}
{Sedov}, L.~I. 1946, Dokl. Akad. Nauk. SSSR, 52, 17

\bibitem[{{Sfeir} {et~al.}(1999){Sfeir}, {Lallement}, {Crifo}, \&
  {Welsh}}]{slcw99}
{Sfeir}, D.~M., {Lallement}, R., {Crifo}, F., \& {Welsh}, B.~Y. 1999, \aap,
  346, 785

\bibitem[{{Shapiro} \& {Field}(1976)}]{sf76}
{Shapiro}, P.~R. \& {Field}, G.~B. 1976, \apj, 205, 762

\bibitem[{{Silich} {et~al.}(2005){Silich}, {Tenorio-Tagle}, \&
  {A{\~n}orve-Zeferino}}]{sta05}
{Silich}, S., {Tenorio-Tagle}, G., \& {A{\~n}orve-Zeferino}, G.~A. 2005, \apj,
  635, 1116

\bibitem[{{Spitoni} {et~al.}(2008){Spitoni}, {Recchi}, \& {Matteucci}}]{s08}
{Spitoni}, E., {Recchi}, S., \& {Matteucci}, F. 2008, \aap, 484, 743

\bibitem[{{Stil} {et~al.}(2009){Stil}, {Wityk}, {Ouyed}, \& {Taylor}}]{s09}
{Stil}, J., {Wityk}, N., {Ouyed}, R., \& {Taylor}, A.~R. 2009, \apj, 701, 330

\bibitem[{{Swinbank}(2007)}]{sw07}
{Swinbank}, M. 2007, in Astronomical Society of the Pacific Conference Series,
  Vol. 379, Cosmic Frontiers, ed. {N.~Metcalfe \& T.~Shanks}, 226

\bibitem[{{Taylor}(1950)}]{t50}
{Taylor}, G. 1950, Royal Society of London Proceedings Series A, 201, 159

\bibitem[{{Tenorio-Tagle} {et~al.}(2003){Tenorio-Tagle}, {Silich}, \&
  {Mu{\~n}oz-Tu{\~n}{\'o}n}}]{tsm03}
{Tenorio-Tagle}, G., {Silich}, S., \& {Mu{\~n}oz-Tu{\~n}{\'o}n}, C. 2003, in
  Revista Mexicana de Astronomia y Astrofisica Conference Series, ed.
  {M.~Reyes-Ruiz \& E.~V{\'a}zquez-Semadeni}, Vol.~18, 136--141

\bibitem[{{Tomisaka} \& {Ikeuchi}(1986)}]{ti86}
{Tomisaka}, K. \& {Ikeuchi}, S. 1986, \pasj, 38, 697

\bibitem[{{T{\"u}llmann} {et~al.}(2006){T{\"u}llmann}, {Pietsch}, {Rossa},
  {Breitschwerdt}, \& {Dettmar}}]{t06}
{T{\"u}llmann}, R., {Pietsch}, W., {Rossa}, J., {Breitschwerdt}, D., \&
  {Dettmar}, R. 2006, \aap, 448, 43

\bibitem[{{Veilleux} {et~al.}(2005){Veilleux}, {Cecil}, \&
  {Bland-Hawthorn}}]{vcb05}
{Veilleux}, S., {Cecil}, G., \& {Bland-Hawthorn}, J. 2005, \araa, 43, 769

\bibitem[{{Weaver} {et~al.}(1977){Weaver}, {McCray}, {Castor}, {Shapiro}, \&
  {Moore}}]{w77}
{Weaver}, R., {McCray}, R., {Castor}, J., {Shapiro}, P., \& {Moore}, R. 1977,
  \apj, 218, 377

\bibitem[{{Welsh} {et~al.}(1999){Welsh}, {Sfeir}, {Sirk}, \&
  {Lallement}}]{wssl99}
{Welsh}, B.~Y., {Sfeir}, D.~M., {Sirk}, M.~M., \& {Lallement}, R. 1999, \aap,
  352, 308

\bibitem[{{West} {et~al.}(2007){West}, {English}, {Normandeau}, \&
  {Landecker}}]{w07}
{West}, J.~L., {English}, J., {Normandeau}, M., \& {Landecker}, T.~L. 2007,
  \apj, 656, 914

\end{thebibliography}
\end{document}